\documentclass[useAMS,usenatbib]{mn2e}
\usepackage{amssymb}
\usepackage{amsmath}
\usepackage{times}
\usepackage{graphicx}
\usepackage{wasysym}

\newcommand{\mincir}{\raise
-2.truept\hbox{\rlap{\hbox{$\sim$}}\raise5.truept 
\hbox{$<$}\ }}
\newcommand{\magcir}{\raise
-2.truept\hbox{\rlap{\hbox{$\sim$}}\raise5.truept
\hbox{$>$}\ }}
\newcommand{\minmag}{\raise-2.truept\hbox{\rlap{\hbox{$<$}}\raise
6.truept\hbox
{$>$}\ }}

\newcommand{\be}{\begin{equation}}
\newcommand{\ee}{\end{equation}}
\newcommand{\ba}{\begin{eqnarray}}
\newcommand{\ea}{\end{eqnarray}}
\newcommand{\brr}{\begin{array}}
 
\newcommand{\err}{\end{array}}
\newcommand{\bc}{\begin{center}}
\newcommand{\ec}{\end{center}}

\newcommand{\tu}{\textunderscore}
\DeclareMathAlphabet{\mathsc}{OT1}{cmr}{m}{sc}
\def\testbx{bx}%
\DeclareRobustCommand{\ion}[2]{%
\relax\ifmmode
\ifx\testbx\f@series
{\mathbf{#1\,\mathsc{#2}}}\else
{\mathrm{#1\,\mathsc{#2}}}\fi
\else\textup{#1\,{\mdseries\textsc{#2}}}%
\fi}

\title[The SFR function of high redshift galaxies]{Simulated star
  formation rate functions at $\bf{z\sim4-7}$, and the role of
  feedback in high-$\bf{z}$ galaxies}

\author[E. Tescari et al.]{E. Tescari$^{1,6}$\thanks{E-mail:
    edoardo.tescari@unimelb.edu.au}, A. Katsianis$^{1,6}$,
  J. S. B. Wyithe$^{1,6}$, K. Dolag$^{2}$, L. Tornatore$^{3,4}$,
  P. Barai$^{3}$, \newauthor{M. Viel$^{3,5}$ and S. Borgani$^{3,4,5}$} \\
  \\ $^1$ School of Physics, University of Melbourne, Parkville, VIC
  3010, Australia \\ $^2$ Universit\"{a}ts-Sternwarte M\"{u}nchen,
  Scheinerstr. 1, D-81679 M\"{u}nchen, Germany \\ $^3$ INAF $-$
  Osservatorio Astronomico di Trieste, Via Tiepolo 11, I-34131
  Trieste, Italy \\ $^4$ Dipartimento di Fisica - Sezione di
  Astronomia, Universit\`{a} di Trieste, Via Tiepolo 11, I-34131 Trieste, Italy \\
  $^5$ INFN $-$ Istituto Nazionale di Fisica Nucleare, Via Valerio 2,
  I-34127 Trieste, Italy \\ $^6$ ARC Centre of Excellence for All-Sky
  Astrophysics (CAASTRO)}

\begin{document}

\maketitle

\begin{abstract}

  We study the role of feedback from supernovae and black holes in the
  evolution of the star formation rate function (SFRF) of $z\sim4-7$
  galaxies. We use a new set of cosmological hydrodynamic simulations,
  {\textsc{Angus}} ({\textit{AustraliaN {\small{GADGET-3}} early
      Universe Simulations}}), run with a modified and improved
  version of the parallel TreePM-smoothed particle hydrodynamics code
  {\small{GADGET-3}} called {\small{P-GADGET3(XXL)}}, that includes a
  self-consistent implementation of stellar evolution and metal
  enrichment. In our simulations both Supernova (SN) driven galactic
  winds and Active Galactic Nuclei (AGN) act simultaneously in a
  complex interplay. The SFRF is insensitive to feedback prescription
  at $z>5$, meaning that it cannot be used to discriminate between
  feedback models during reionisation. However, the SFRF is sensitive
  to the details of feedback prescription at lower redshift. By
  exploring different SN driven wind velocities and regimes for the
  AGN feedback, we find that the key factor for reproducing the
  observed SFRFs is a combination of ``strong'' SN winds and early AGN
  feedback in low mass galaxies. Conversely, we show that the choice
  of initial mass function and inclusion of metal cooling have less
  impact on the evolution of the SFRF. When variable winds are
  considered, we find that a non-aggressive wind scaling is needed to
  reproduce the SFRFs at $z\apprge4$. Otherwise, the amount of objects
  with low SFRs is greatly suppressed and at the same time winds are
  not effective enough in the most massive systems.

\end{abstract}


\section{Introduction}

A detailed understanding of galaxy formation and evolution in the cold
dark matter model remains a challenge for modern cosmology, especially
at high redshift. The star formation history is determined by a
complex interplay of gas accretion onto the potential wells created by
the Dark Matter (DM) halos, star formation and associated feedback
processes. Cosmological simulations have become a powerful tool to
study how these astrophysical phenomena interact and influence each
other \citep{schaye10,finlator11,dave11,maio11,choi12,jaacks12,
  puchespring13,wilkins13,kannan13,Vogel13}. However, despite the
incredible growth of available computating resources seen in recent
years (in terms of both hardware and software), numerical codes that
follow both dark matter and baryonic physics still have shortcomings,
mainly because of the complexity and the different range of scales of
the physics involved. It is, for example, numerically infeasible to
fully resolve all the relevant processes that govern the formation of
stars in galaxies and their backreaction on the interstellar medium
\citep{puchespring13}. For this reason, current cosmological
simulations use sub-resolution schemes to model the physics of star
formation, supernova explosions and black hole gas accretion. On the
other hand, numerical techniques have been substantially improved
during the last few years \citep[e.g.][]{Murante11, ReadHayfield12,
  Hopkins13}. Particularly interesting is the moving mesh approach,
since it combines the best features of grid and particle based codes
\citep{Springel10,Sijacki12,Keres12,Vogel12,Torrey12,Nelson13,Vogel13}.

Observationally, our knowledge of Star Formation Rates (SFRs) and
stellar masses in the distant Universe has expanded substantially over
the last several years thanks to deep multi-wavelength surveys
\citep{bernardi10,gonzalez11,Lee2012,smit12,santini12,bouwens2012,
  schenker12}. In particular, the evolution of the star formation rate
over cosmic time is a fundamental constraint for theories of galaxy
formation and evolution
\citep{hopkins04,wilkins08,Guo11,cucciati12,bouwens2012}. Furthermore,
Star Formation Rate Functions (SFRFs) provide a physical description
of galaxy buildup at high redshift. In a recent work, \citet{smit12}
combined the UV luminosity functions of
\citet{bouwens2007,bouwens2011} with new estimates of dust extinction,
to derive the SFRF of $z\sim 4-7$ galaxies. Their results provide
strong indications that galaxies build up uniformly over the first 3
Gyr of the Universe.

The star formation rate functions at lower redshift have been studied
by means of both simulations \citep{dave11} and semi-analytic models
\citep{fontanot12}. In particular, \citet{dave11} used a set of
simulations run with an extended version of the hydrodynamic code
{\small{GADGET-2}} to study the growth of the stellar content of
galaxies at redshift $0\le z\le3$. The authors implemented four
different wind models and simulated stellar mass and star formation
rate functions to quantify the effects of outflows on galactic
evolution. In their simulations, galactic winds are responsible for
the shape of the faint end slope of the SFR function at
$z=0$. However, feedback from Active Galactic Nuclei (AGN) is not
taken into account and, as a result, their simulations overproduce the
number of objects in the high star formation rate
tail. \citet{puchespring13}, instead, used {\small{GADGET-3}} to
investigate how both supernova and AGN feedback processes affect the
shape of the galaxy stellar mass functions (GSMFs) at redshift $0\le
z\le 2$. These authors suggest that an energy-driven wind model in
which the wind velocity decreases and the wind mass loading increases
in low mass galaxies \citep{Martin2005}, can produce a good match to
the low mass end of the observed GSMF. Conversely, the high mass end
can be recovered simultaneously if AGN feedback is included. The
importance of AGN feedback in shaping the high end of the luminosity
functions at low redshift is also indicated by semi-analytic models
\citep{croton06,bower06}.

The aim of this work is to investigate the relative importance of
galactic winds and AGN feedback at higher redshift. We use
state-of-the-art cosmological hydrodynamic simulations run with a
modified and improved version of the widely used Smoothed Particle
Hydrodynamics (SPH) code {\small{GADGET-3}} \citep[last described
in][]{springel2005} called {\small{P-GADGET3(XXL)}}. As a first
scientific case, in this paper we study the evolution of the star
formation rate function of high redshift galaxies. In a companion
paper \citep{kata13} we will explore the stellar mass function and the
star formation rate$-$stellar mass relation for high redshift
galaxies.

The paper is organized as follows. In Sections \ref{thecode} and
\ref{angus} we present the hydrodynamic code {\small{P-GADGET3(XXL)}}
and our simulation project {\textsc{Angus}}. In Section \ref{feedb} we
discuss the different feedback models included in our runs and in
Section \ref{simout} we describe the characteristics of the
simulations used in this work. In Section \ref{cosmsfr} we study the
evolution of the cosmic star formation rate density. In Section
\ref{obs_sfrf} we summarise observations of the star formation rate
function of high redshift galaxies from \citet{smit12}, and compare
with our simulations in Section \ref{sim_sfrf}. In Section
\ref{discussion} we discuss our main results, and present conclusions
in Section \ref{concl}.

\section{The simulations}
\label{thesims}

The set of simulations used in this work is part of the
{\textit{AustraliaN {\small{GADGET-3}} early Universe Simulations}}
({\textsc{Angus}}) project. We have started this project to study the
interplay between galaxies and the Intergalactic Medium (IGM) from low
redshift ($z<2$) to the epoch of reionisation at $z\sim6$ and above.

\subsection{The code}
\label{thecode}

{\small{P-GADGET3(XXL)}} (PG3) is an improved version of
{\small{GADGET-3}} \citep{springel2005}. For the first time it
combines a number of physical processes, which have been developed and
tested separately. In particular, for this work we used:
\begin{itemize}
\item a sub-grid star formation model \citep{springel2003};
\item self-consistent stellar evolution and chemical enrichment
  modeling \citep{T07};
\item supernova energy- and momentum-driven galactic winds
  \citep{springel2003,puchespring13};
\item AGN feedback \citep{springeletal05,dunja2010,susana13};
\item metal-line cooling \citep{wiersma09};
\item a low-viscosity SPH scheme to allow the development of
  turbulence within the intracluster medium \citep{dolag2005b}.
\end{itemize}
Additional physical modules will be explored in the future:
\begin{itemize}
\item new supernova driven galactic wind feedback prescriptions
  \citep{barai13};
\item transition of metal free Population III to Population II star
  formation \citep{tornatore07b}; %change here file angus_5r.bbl
\item low-temperature cooling by molecules/metals \citep{maio2007};
\item adaptive gravitational softening \citep{iannuzzi11};
\item thermal conduction \citep{dolag2004};
\item passive magnetic fields based on Euler potentials
  \citep{dolagstasy09}.
\end{itemize}
Moreover, the following on the fly tools have been added to PG3:
\begin{itemize}
\item parallel Friends-of-Friends (FoF) algorithm to identify
  collapsed structures. The FoF links over all particle types (dark
  matter, gas and stars), and enables combinations of one, two or
  three types. Following \citet{klaus_subf09}, we use a linking length
  of 0.16 times the mean DM particle separation\footnote{This linking
    length is obtained by re-scaling the standard linking length of
    0.2 according to the adopted $\Lambda$CDM cosmology (see Section
    \ref{angus}).};
\item parallel {\small{SUBFIND}} algorithm to identify substructures
  within FoF halos \citep{subfind,klaus_subf09}. The formation of
  purely gaseous substructures is prevented and, if the
  \citet{chabrier03} Initial Mass Function (IMF) is used,
  {\small{SUBFIND}} can assign luminosities in 12 bands (u, V, G, r,
  i, z, Y, J, H, K, L, M) to subhalos, including dust attenuation
  (based on spectral energy distributions from Charlot \& Bruzual in
  preparation).
\end{itemize}

PG3 is currently being used for a range of different scientific
projects, including study of galaxy clusters and magnetic fields
within the large scale structures.

\subsection{The {\textsc{ANGUS}} project}
\label{angus}

%\begin{table*}
%\centering
%\begin{tabular}{llccccc}
%  \\ \hline & Name & Box size & Number of & M$_{\rm DM}$  & M$_{\rm
%    GAS}$ & Comoving softening \\ &  & [Mpc/$h$] & particles & [M$_{\rm
%    \odot}$/$h$] & [M$_{\rm \odot}$/$h$] & [kpc/$h$] \\ \hline
  % & Box4$_{\rm mr}$ & 48 & 2$\times81^3$ & 1.31$\times10^{10}$ & &
  % 30.0 \\
%  & Box4a$_{\rm uhr}$ & 24 & 2$\times288^3$ & 3.64$\times10^{7}$ & 7.32$\times10^{6}$ & 4.0 \\ 
%  & Box5$_{\rm hr}$ & 18 & 2$\times81^3$ & 6.90$\times10^{8}$ & 1.39$\times10^{8}$ & 10.0 \\ 
%  & Box5$_{\rm uhr}$ & 18 & 2$\times216^3$ & 3.64$\times10^{7}$ & 7.32$\times10^{6}$ & 4.0 \\ 
%  & Box5$_{\rm axhr}$ & 18 & 2$\times384^3$ & 6.47$\times10^{6}$ & 1.30$\times10^{6}$ &
%  2.0 \\ 
%  & Box5$_{\rm xhr}$ & 18 & 2$\times576^3$ & 1.92$\times10^{6}$ & 3.86$\times10^{5}$ & 1.5 \\ 
%  & Box6$_{\rm uhr}$ & 12 & 2$\times144^3$ & 3.64$\times10^{7}$ & 7.32$\times10^{6}$ & 4.0 \\ 
%  & Box6$_{\rm axhr}$ & 12 & 2$\times256^3$ & 6.47$\times10^{6}$ & 1.30$\times10^{6}$ & 2.0 \\ 
%  & Box6$_{\rm xhr}$ & 12 & 2$\times384^3$ & 1.92$\times10^{6}$ & 3.86$\times10^{5}$ & 1.5 \\ 
%  \hline \\
%\end{tabular}
%\caption{Configurations of the {\textsc{Angus}} project. Columns are as follows:
%  (1) name: $hr=$ high resolution, $uhr=$
%  ultra-high res, $axhr=$ almost-extremely high res and
%  $xhr=$ extremely high res; (2) box size; (3) total number of
%  particles (N$_{\rm GAS}$ $+$ N$_{\rm DM}$); (4) mass of the dark matter
%  particles; (5) initial mass of the gas particles; (6)
%  Plummer-equivalent comoving gravitational softening length.} 
%\label{tab:ang_conf}
%\end{table*}

We have performed a set of cosmological simulations assuming a flat
$\Lambda$CDM model with $\Omega_{\rm 0m}=0.272$, $\Omega_{\rm
  0b}=0.0456$, $\Omega_{\rm \Lambda}=0.728$, $n_{\rm s}=0.963$,
$H_{\rm 0}=70.4$ km s$^{-1}$ Mpc$^{-1}$ (or $h=0.704$) and
$\sigma_{\rm 8}=0.809$. This set of cosmological parameters is the
combination of 7-year data from WMAP \citep{komatsu11} with the
distance measurements from the baryon acoustic oscillations in the
distribution of galaxies \citep{percival10} and the Hubble constant
measurement of \citet{riess09}\footnote{Note that some of these
  parameters are in tension with recent results from the
  \textit{Planck} satellite \citep{Planck13}.}.
% The various simulation configurations for the {\textsc{Angus}}
% project are listed in Table \ref{tab:ang_conf}. In this work we use
% only the Box4a$_{\rm uhr}$, Box5$_{\rm axhr}$ and Box6$_{\rm xhr}$
% configurations, run in different versions (see Section
% \ref{simout}).

Each simulation produces a cosmological box with periodic boundary
conditions and initially contains an equal number of gas and dark
matter particles. We adopt the multiphase star formation criterion of
\citet{springel2003}, where a prescription for the Interstellar Medium
(ISM) is included. In this model, the ISM is represented as a fluid
comprised of cold condensed clouds in pressure equilibrium with an
ambient hot gas. The clouds supply the material available for star
formation. Whenever the density of a gas particle exceeds a given
threshold $\rho_{\rm th}$, that gas particle is flagged as star
forming and is treated as multiphase. With this prescription, baryons
are in the form of either a hot or a cold phase or in stars, so that
this density threshold marks the onset of cold cloud formation. Mass
exchange between the different phases is driven by the effect of star
formation, cloud evaporation and cooling. A typical value for
$\rho_{\rm th}$ is $\sim0.1$ cm$^{-3}$ (in terms of the number density
of hydrogen atoms), but the exact density threshold is calculated
according to the IMF used and the inclusion/exclusion of metal-line
cooling (see below). This guarantees that our simulations generate the
Kennicutt-Schmidt law \citep{Kenni1998} by construction.

PG3 self-consistently follows the evolution of hydrogen, helium and 9
metallic species (C, Ca, O, N, Ne, Mg, S, Si and Fe) released from
Supernovae (SNIa and SNII) and Low and Intermediate Mass Stars
(LIMSs). Radiative cooling and heating processes are included
following the procedure presented in \citet{wiersma09}. We assume a
mean background radiation composed of the cosmic microwave background
and the \citet{haardtmadau01} ultraviolet/X-ray background from
quasars and galaxies. Contributions to cooling from each of the eleven
elements mentioned above have been pre–computed using the {\small
  {Cloudy}} photo–ionisation code \citep[last described
in][]{ferland13} for an optically thin gas in (photo)ionisation
equilibrium. In this work we use cooling tables for gas of primordial
composition (H $+$ He) as the reference configuration. To test the
effect of metal-line cooling, we include it in one simulation. It is
worth noting that this procedure for computing cooling rates allows us
to relax the assumptions of collisional ionisation equilibrium and
solar relative abundances of the elements previously adopted
\citep{T07,tex09,tex11}. Mixing and diffusion are not included in our
simulations. However, to make sure that metals ejected by stars in the
ISM mix properly, for each gas particle in the ISM we consider a
``smoothed metallicity'' calculated taking into account the
contribution of neighbouring SPH particles.  This procedure
effectively reduces the noise associated with the lack of diffusion
and mixing.

Our model of chemical evolution accounts for the age of various
stellar populations and metals, which are released over different time
scales by stars of different mass. We adopt the lifetime function of
\citet{padovanimatteucci93} and the following stellar yields:
\begin{itemize}
\item SNIa: \citet{thielemann03}. The mass range for the SNIa
  originating from binary systems is 0.8 M$_{\rm \odot}<m\le 8$
  M$_{\rm \odot}$, with a binary fraction of 7 per cent.
\item SNII: \citet{woosleyweaver95}. The mass range for the SNII is
  $m>8$ M$_{\rm \odot}$.
\item LIMS: \citet{vandenhoek97}.
\end{itemize}
In the code, every star particle represents a Simple Stellar
Population (SSP). To describe this SSP, we take into account stars
with mass in the interval 0.1 M$_{\rm \odot}\le m\le100$ M$_{\rm
  \odot}$. Only stars with mass $m\le40$ M$_{\rm \odot}$ explode as SN
before turning into black holes, while stars of mass $m>40$
M$_{\rm \odot}$ collapse directly to a black hole without forming a SN. \\

An important ingredient of the model is the initial stellar mass
function $\xi(m)$, which defines the distribution of stellar masses
(dN) per logarithmic mass interval that form in one star formation
event in a given volume of space:
\begin{eqnarray}
  \xi(m) = \frac{{\rm d}\,{\rm N}}{{\rm d} \log m}.
\end{eqnarray}
The most widely used functional form for the IMF is the power law, as
suggested originally by \citet{salpeter55}:
\begin{eqnarray}
  \xi(m) = A\,m^{-x},
\end{eqnarray}
where $A$ is a normalization factor, set by the condition:
\begin{eqnarray}
  \int_{m_{\rm inf}}^{m_{\rm sup}}\xi(m)\,{\rm d}m=1,
\end{eqnarray}
where in our case $m_{\rm inf}=0.1$ M$_{\rm \odot}$ and $m_{\rm
  sup}=100$ M$_{\rm \odot}$. In this work we use three different
initial stellar mass functions:
\begin{itemize}
\item \citet{salpeter55}: single sloped,
\begin{eqnarray}
  \xi(m)=0.172\times m^{-1.35}
\end{eqnarray}
\item \citet{kroupa93}: multi sloped,
\begin{eqnarray}
  \xi(m)=\left\{\begin{array}{l}0.579\times m^{-0.3}\ \ 0.1\,{\rm M}_{\rm
        \odot}\le m <0.5\,{\rm M}_{\rm
        \odot}\\0.310\times m^{-1.2} \ \ 0.5\,{\rm M}_{\rm
        \odot}\le m<1\,{\rm M}_{\rm
        \odot}\\0.310\times m^{-1.7}\ \ m\ge1\,{\rm M}_{\rm
        \odot}
\end{array}\right.
\end{eqnarray}
\item \citet{chabrier03}: multi sloped\footnote{Note that the original
    \citet{chabrier03} IMF is a power-law for stellar masses $m\ge1$
    M$_{\rm \odot}$ and has a log-normal form at lower
    masses. However, since our code can easily handle multi sloped
    IMFs, we use a power law approximation of the \citet{chabrier03}
    IMF. It consists of three different slopes over the whole mass
    range $0.1-100$ M$_{\rm \odot}$, to mimic the log-normal form of
    the original one at low masses.},
\begin{eqnarray}
  \xi(m)=\left\{\begin{array}{l}0.497\times m^{-0.2}\ \ 0.1\,{\rm M}_{\rm
        \odot}\le m< 0.3\,{\rm M}_{\rm
        \odot}\\0.241\times m^{-0.8} \ \ 0.3\,{\rm M}_{\rm
        \odot}\le m<1\,{\rm M}_{\rm 
        \odot}\\0.241\times m^{-1.3}\ \ m\ge1\,{\rm M}_{\rm 
        \odot}\end{array}\right.
\end{eqnarray}
\end{itemize}

%\begin{figure}
%\centering 
%\includegraphics[width=8.5cm]{imfs.ps}
%\caption{Initial stellar mass functions used in this work:
%  \citet[][red solid line]{salpeter55}, \citet[][blue dashed
%  line]{chabrier03} and \citet[][black dot-dashed line]{kroupa93}.}
%\label{fig_imfs}
%\end{figure}

The initial mass function is particularly important, because the
stellar mass distribution determines the evolution, surface
brightness, chemical enrichment, and baryonic content of galaxies. The
shape of the IMF also determines how many long-lived stars form with
respect to massive short-lived stars. In turn, this ratio affects the
amount of energy released by SN and the current luminosity of galaxies
which is dominated by low-mass stars.
% In Figure \ref{fig_imfs} we compare the three IMFs.
The Salpeter IMF predicts a larger number of low-mass stars, while the
Kroupa and the Chabrier IMFs predict a larger number of intermediate
and high-mass stars, respectively. However, note that we neglect the
effect of assuming different IMFs on the observationally inferred
cosmic SFR. Therefore, the choice of the IMF has an indirect (minor)
impact on the evolution of the cosmic star formation rate density and
on the star formation rate functions, as we discuss in Section
\ref{cosmsfr} and \ref{sim_sfrf}, respectively. Conversely, the same
choice produces different enrichment patterns as shown in \citet{T07}.

\subsection{Feedback models}
\label{feedb}

In our simulations both SN driven galactic winds and AGN feedback act
simultaneously in a complex interplay. In one run we adopt variable
momentum-driven galactic winds, while all the other simulations
include energy-driven galactic winds of constant velocity. In the
following sub-sections we discuss our different feedback models.

\subsubsection{Energy-driven galactic winds}

We use the \citet{springel2003} implementation of energy-driven
galactic winds. The wind mass-loss rate ($\dot{\rm M}_{\rm w}$) is
assumed to be proportional to the star formation rate ($\dot{\rm
  M}_{\rm \star}$) according to:
\begin{eqnarray}
  \dot{\rm M}_{\rm w} = \eta\dot{\rm M}_{\rm \star},
  \label{eq_windload}
\end{eqnarray}
where the wind mass loading factor $\eta$ is a parameter that accounts
for the wind efficiency. The wind carries a fixed fraction $\chi$ of
the SN energy $\epsilon_{\rm SN}$:
\begin{eqnarray}
  \frac{1}{2}\dot{\rm M}_{\rm w}v_{\rm w}^2 = \chi\,\epsilon_{\rm
    SN}\dot{\rm M}_{\rm \star}\,\,\,\,\Longrightarrow\,\,\,\,v_{\rm w} = \sqrt{\frac{2\,\chi\,\epsilon_{\rm
        SN}}{\eta}},
\end{eqnarray}
where $v_{\rm w}$ is the wind velocity. Star-forming gas particles are
then stochastically selected according to their SFR to become part of
a blowing wind. Whenever a particle is uploaded to the wind, it is
decoupled from the hydrodynamics for a given period of time. The
maximum allowed time for a wind particle to stay hydrodynamically
decoupled is $t_{\rm dec} = l_{\rm w}/v_{\rm w}$, where $l_{\rm w}$ is
the wind free travel length. In addition, when $t<t_{\rm dec}$, a wind
particle will re-couple to the hydrodynamics as soon as it reaches a
region with a density:
\begin{eqnarray}
  \rho<\delta_{\rm w}\rho_{\rm th},
\end{eqnarray}
where $\rho_{\rm th}$ is the density threshold for the
onset of the star formation and $\delta_{\rm w}$ is the wind free
travel density factor. The parameters $t_{\rm dec}$ and $\delta_{\rm
  w}$ have been introduced in order to prevent a gas particle from
getting trapped into the potential well of the virialized halo, thus
allowing effective escape from the ISM into the low-density IGM.

We consider the velocity of the wind $v_{\rm w}$ as a free parameter
\citep{tornatore010}. The following additional parameters therefore
fully specify the wind model: $\eta=2$ \citep{martin1999}, $t_{\rm
  dec}=0.025\left[\frac{{\rm comoving\ kpc}/h}{{\rm km/s}}\right]$
(code internal units corresponding to $\sim34.7$ Myr) and
$\delta_{\rm w}=0.5$ ($\chi$ being fixed by the choice of $v_{\rm
  w}$). In this work we explore three different values of $v_{\rm w}$:
\begin{itemize}
\item weak winds: $v_{\rm w}=350$ km/s. Since in code internal units
  $t_{\rm dec} = l_{\rm w}/v_{\rm w}=0.025\left[\frac{{\rm comoving\
        kpc}/h}{{\rm km/s}}\right]$, in this case the wind free travel
  length is $l_{\rm w}=8.75$ kpc/$h$ (comoving);
\item strong winds: $v_{\rm w}=450$ km/s ($l_{\rm w}=11.25$ kpc/$h$);
\item very strong winds: $v_{\rm w}=550$ km/s ($l_{\rm w}=13.75$
  kpc/$h$).
\end{itemize}

\subsubsection{Momentum-driven galactic winds}

To test the sensitivity of our conclusions to the adopted wind model,
we performed a simulation that includes variable winds. According to
the observational results of \citet{Martin2005}, we assume that in a
given halo the velocity of the wind depends on the mass of the halo
M$_{\rm halo}$:
\begin{eqnarray}
  v_{\rm w}= 2\;\sqrt{\frac{G{\rm M}_{\rm halo}}{R_{\rm 200}}}=2\times v_{\rm circ},
\end{eqnarray}
where $v_{\rm circ}$ is the circular velocity and $R_{\rm 200}$ is the
radius within which a density 200 times the mean density of the
Universe at redshift $z$ is enclosed \citep{barai13}:
%\begin{eqnarray}
%      %  M_{\rm halo} = \frac{4\pi}{3}\left(200\rho_{\rm
%      c}\Omega_{\rm 0m}\right)\left(1+z\right)^3.
%\end{eqnarray}
\begin{eqnarray}
  R_{\rm 200} = \sqrt[3]{\frac{3}{4\pi}\frac{{\rm M}_{\rm halo}}{200\rho_{\rm
        c}\Omega_{\rm 0m}}}\left(1+z\right)^{-1}.
\end{eqnarray}
In this equation $\rho_{\rm c}=3H_{\rm 0}^2/(8\pi G)$ is the critical
density at $z=0$. Following \citet{puchespring13}, we consider a
momentum-driven scaling of the wind mass loading factor:
\begin{eqnarray}
  \eta = 2\times\frac{450\,\,{\rm km/s}}{v_{\rm w}},
\end{eqnarray}
where $\eta=2$ if the wind velocity is equal to our reference constant
(strong) wind model $v_{\rm w}=450$ km/s. Here again, we decouple wind
particles from the hydrodynamics for a given period of time. \\

Besides the kinetic (energy- or momentum-driven) feedback just
described, contributions from both SNIa and SNII to thermal feedback
are also considered.

\begin{table*}
\centering
\begin{tabular}{llccccccc}
  \\ \hline & Run & IMF & Box Size & N$_{\rm TOT}$ & M$_{\rm DM}$ & M$_{\rm GAS}$  &
  Comoving Softening & Feedback \\ & & & [Mpc/$h$] &
  & [M$_{\rm \odot}$/$h$] & [M$_{\rm \odot}$/$h$] & [kpc/$h$] \\ \hline
  & \textit{Kr24\tu eA\tu sW} & Kroupa & 24 & $2\times288^3$ &
  3.64$\times10^{7}$ & $7.32\times 10^6$ & 4.0 & Early AGN $+$ Strong Winds \\
  & \textit{Ch24\tu lA\tu wW} & Chabrier & 24 & $2\times288^3$ &
  3.64$\times10^{7}$ & $7.32\times 10^6$ & 4.0 & Late AGN $+$ Weak Winds \\
  & \textit{Sa24\tu eA\tu wW} & Salpeter & 24 & $2\times288^3$ &
  3.64$\times10^{7}$ & $7.32\times 10^6$ & 4.0 & Early AGN $+$ Weak Winds \\
  & \textit{Ch24\tu eA\tu sW} & Chabrier & 24 & $2\times288^3$ & 3.64$\times10^{7}$ & $7.32\times 10^6$ & 4.0 & Early AGN $+$ Strong Winds  \\
  & \textit{Ch24\tu lA\tu sW} & Chabrier & 24 & $2\times288^3$ &
  3.64$\times10^{7}$ & $7.32\times 10^6$ & 4.0 & Late AGN $+$ Strong Winds \\
  & \textit{Ch24\tu eA\tu vsW} & Chabrier & 24 & $2\times288^3$ &
  3.64$\times10^{7}$ & $7.32\times 10^6$ & 4.0 & Early AGN $+$ Very Strong Winds \\
  & \textit{Ch24\tu NF} & Chabrier & 24 & $2\times288^3$ &
  3.64$\times10^{7}$ & $7.32\times 10^6$ & 4.0 & No Feedback\\
  & \textit{Ch24\tu Zc\tu eA\tu sW}$^a$ & Chabrier & 24 &
  $2\times288^3$ & 3.64$\times10^{7}$ & 
  $7.32\times 10^6$ & 4.0 & Early AGN $+$ Strong Winds \\
  & \textit{Ch24\tu eA\tu MDW}$^b$ & Chabrier & 24 & $2\times288^3$ &
  3.64$\times10^{7}$ & $7.32\times 10^6$ & 4.0 & Early AGN $+$ \\
  & & & & & & & & Momentum-Driven Winds \\
  & \textit{Ch18\tu lA\tu wW} & Chabrier & 18 & $2\times384^3$ &
  6.47$\times10^{6}$ & $1.30\times 10^6$ & 2.0 & Late AGN $+$ Weak Winds \\
  & \textit{Ch12\tu eA\tu sW} & Chabrier & 12 & $2\times384^3$ &
  1.92$\times10^{6}$ & $3.86\times 10^5$ & 1.5 & Early AGN $+$ Strong Winds \\
  \hline \\
\end{tabular}
\caption{Summary of the different runs used in this work. Column 1, run name; column 2,
  Initial Mass Function (IMF) chosen; column 3, box size in comoving Mpc/$h$;
  column 4, total number of particles (N$_{\rm TOT} =$ N$_{\rm
    GAS}$ $+$ N$_{\rm DM}$); column 5, mass of the dark matter particles; column 6, initial mass of the gas particles;
  column 7, Plummer-equivalent comoving gravitational softening length; column 8, type of feedback
  implemented. See Section \ref{simout} for
  more details on the parameters used for the different feedback
  recipes. $(a)$: in this simulation the effect of metal-line cooling is
  included (Section \ref{angus}). $(b)$: in this simulation we
  adopt variable momentum-driven galactic winds. In all the other simulations
  galactic winds are energy-driven and the wind velocity is constant. (Section \ref{feedb}).}
\label{tab:sim_runs}
\end{table*}

\subsubsection{AGN feedback}
\label{agn_feedb}

Our model for AGN feedback is based on the implementation proposed by
\citet{springeletal05}, with feedback energy released from gas
accretion onto Super-Massive Black Holes (SMBHs)\footnote{See
  \citet{boothschaye09} for a different implementation and
  \citet{wurster13} for a comparative study of AGN feedback
  algorithms.}. However, our model introduces some modifications
\citep{dunja2010,susana13}. SMBHs are described as collisionless sink
particles initially seeded in dark matter halos, which grow via gas
accretion and through mergers with other SMBHs during close
encounters. Whenever a dark matter halo, identified by the parallel
run-time FoF algorithm, reaches a mass above a given mass threshold
M$_{\rm th}$ for the first time, it is seeded with a central SMBH of
mass M$_{\rm seed}$. Each SMBH can then grow by accreting local gas at
a Bondi rate (Eddington-limited):
\begin{eqnarray}
  \dot{\rm M}_{\rm SMBH}={\rm min}\left(\dot{\rm M}_{\rm B}, \dot{\rm M}_{\rm
      Edd}\right),
\label{eq:acrate}
\end{eqnarray}
where $\dot{\rm M}_{\rm Edd}$ and $\dot{\rm M}_{\rm B}$ are the
Eddington and the Bondi \citep{bondi1952} accretion rates,
respectively. According to Eq. (\ref{eq:acrate}), the theoretical
accretion rate value is computed throughout the evolution of the
simulation for each SMBH. We numerically implement this accretion rate
using a stochastic criterion to decide which of the surrounding gas
particles contribute to the accretion. In the model of
\citet{springeletal05}, a selected gas particle contributes to
accretion with all its mass. Our model allows for a gas particle to
supply only 1/4 of its original mass. As a result, each gas particle
can contribute to up to four generations of SMBH accretion events. In
this way, a larger number of particles are involved in the accretion,
which is then followed in a more continuous way.

The radiated energy in units of the energy associated with the
accreted mass is:
\begin{eqnarray}
  L_{\rm r}=\epsilon_{\rm r}\dot{\rm M}_{\rm SMBH}\,c^2\,\,\,\,\Longrightarrow\,\,\,\,\epsilon_{\rm r}=\frac{L_{\rm r}}{\dot{\rm M}_{\rm SMBH}\,c^2},
\end{eqnarray}
where $\epsilon_{\rm r}$ is the radiative efficiency of the SMBH. The
model assumes that a fraction $\epsilon_{\rm f}$ of the radiated
energy is thermally coupled to the surrounding gas according to:
\begin{eqnarray}
  \dot{E}_{\rm feed}=\epsilon_{\rm f}\epsilon_{\rm r}\dot{\rm M}_{\rm SMBH}\,c^2.
\end{eqnarray}
We set $\epsilon_{\rm r} = 0.1$, coincident with the mean
$\epsilon_{\rm r}$ value for radiatively efficient accretion onto a
Schwarzschild black hole (\citealp{shakura1973}; see also
\citealp{maio13}).
% The SMBH mass is correspondingly decreased by this amount.
At high redshift, SMBHs are characterized by high accretion rates and
power very luminous quasars, with only a small fraction of the
radiated energy being thermally coupled to the surrounding gas. For
this reason, following \citet{Wyithe03}, \citet{sizicki07} and
\citet{dunja2010}, we use $\epsilon_{\rm f} = 0.05$. On the other
hand, SMBHs hosted within very massive halos at lower redshift are
expected to accrete at a rate well below the Eddington limit, while
the energy is mostly released in a kinetic form, and eventually
thermalized in the surrounding gas through shocks.  Therefore,
whenever accretion enters in the quiescent radio mode and takes place
at a rate smaller than one-hundredth of the Eddington limit we
increase the feedback efficiency to $\epsilon_{\rm f} = 0.2$.

Some other technical details distinguish our model for AGN
feedback. In order to guarantee that SMBHs are seeded only in halos
where substantial star formation took place, we impose the condition
that such halos should contain a minimum mass fraction in stars
$f_{\star}$, with only halos having M$_{\star}\ge f_{\star}\times{\rm
  M}_{\rm th}$ being seeded with a black hole. Furthermore, we locate
seeded SMBHs at the potential minimum of the FoF group, instead of at
the density maximum, as implemented by \citet{springeletal05}. We also
enforce a strict momentum conservation during gas accretion and SMBH
mergers. In this way a SMBH particle remains within the host galaxy
when it becomes a satellite of a larger halo.

In this work we consider two regimes for AGN feedback, where we vary
the minimum FoF mass M$_{\rm th}$ and the minimum star mass fraction
$f_{\star}$ for seeding a SMBH, the mass of the seed M$_{\rm seed}$
and the maximum accretion radius $R_{\rm ac}$.
% In analogy with the different wind recipes presented above,
We define:
\begin{itemize}
\item early AGN formation: M$_{\rm th}=2.9\times10^{10}$ M$_{\rm
    \odot}/h$, $f_{\star}=2.0\times10^{-4}$, M$_{\rm seed}=5.8\times
  10^{4}$ M$_{\rm \odot}/h$, $R_{\rm ac}=200$ kpc/$h$;
\item late AGN formation: M$_{\rm th}=5.0\times10^{12}$ M$_{\rm
    \odot}/h$, $f_{\star}=2.0\times10^{-2}$, M$_{\rm seed}=2.0\times
  10^{6}$ M$_{\rm \odot}/h$, $R_{\rm ac}=100$ kpc/$h$.
\end{itemize}
We stress that the radiative efficiency ($\epsilon_{\rm r}$) and the
feedback efficiency ($\epsilon_{\rm f}$) are assumed to be the same in
the two regimes. However, in the early AGN configuration we allow the
presence of a black hole in lower mass halos, and at earlier
times. Because very large star forming galaxies are very rare in our
high redshift simulations, the late AGN scheme includes almost no AGN
feedback in the regime we consider. In the early AGN case, M$_{\rm
  seed}/\left(f_{\star}\times{\rm M}_{\rm th}\right)=10^{-2}$, an
order of magnitude larger than the local Magorrian relation
\citep{magorrian1998}, leading to significant feedback in low mass
galaxies at high redshift.

\subsection{Outline of simulations}
\label{simout}

In Table \ref{tab:sim_runs} we summarise the main parameters of the
cosmological simulations performed for this work. Our reference
configuration has box size $L=24$ Mpc/$h$, initial mass of the gas
particles M$_{\rm GAS}=7.32\times 10^6$ M$_{\rm \odot}/h$ and a total
number of particles (N$_{\rm TOT} =$ N$_{\rm GAS}$ $+$ N$_{\rm DM}$)
equal to $2\times288^3$. We also ran two simulations with $L=18$
Mpc/$h$ and $L=12$ Mpc/$h$ to perform box size and resolution
tests. All the simulations start at $z=60$ and were stopped at
$z=2$. In the following we outline the characteristics of each run:
\begin{itemize}
\item {\bf{\textit {Kr24\tu eA\tu sW}}}: \citet{kroupa93} initial mass
  function, box size $L=24$ Mpc/$h$, early AGN feedback and strong
  energy-driven galactic winds of velocity $v_{\rm w}=450$ km/s;
\item {\bf{\textit {Ch24\tu lA\tu wW}}}: \citet{chabrier03} IMF, late
  AGN feedback and weak winds with $v_{\rm w}=350$ km/s;
\item {\bf{\textit {Sa24\tu eA\tu wW}}}: \citet{salpeter55} IMF,
  early AGN feedback and weak winds with $v_{\rm w}=350$ km/s;
\item {\bf{\textit {Ch24\tu eA\tu sW}}}: Chabrier IMF, early AGN
  feedback and strong winds with $v_{\rm w}=450$ km/s;
\item {\bf{\textit {Ch24\tu lA\tu sW}}}: Chabrier IMF, late AGN
  feedback and strong winds with $v_{\rm w}=450$ km/s;
\item {\bf{\textit {Ch24\tu eA\tu vsW}}}: Chabrier IMF, early AGN
  feedback and very strong winds with $v_{\rm w}=550$ km/s;
\item {\bf{\textit {Ch24\tu NF}}}: Chabrier IMF. This simulation was
  run without any winds or AGN feedback, in order to test how large
  the effects of the different feedback prescriptions are;
\item {\bf{\textit {Ch24\tu Zc\tu eA\tu sW}}}: Chabrier IMF, metal
  cooling, early AGN feedback and strong winds with $v_{\rm w}=450$
  km/s;
\item {\bf{\textit {Ch24\tu eA\tu MDW}}}: Chabrier IMF, early AGN
  feedback and momentum-driven galactic winds. For the wind mass
  loading factor we used the same scaling of Eq. (4) in
  \citet{puchespring13};
\item {\bf{\textit {Ch18\tu lA\tu wW}}}: Chabrier IMF, box size $L=18$
  Mpc/$h$, late AGN feedback and weak winds of velocity $v_{\rm
    w}=350$ km/s. The initial mass of the gas particles is M$_{\rm
    GAS}=1.30\times10^{6}$ M$_{\rm \odot}/h$ and the total number of
  particles is equal to $2\times384^3$;
\item {\bf{\textit {Ch12\tu eA\tu sW}}}: Chabrier IMF, box size $L=12$
  Mpc/$h$, early AGN feedback and strong winds of velocity $v_{\rm
    w}=450$ km/s. The initial mass of the gas particles is M$_{\rm
    GAS}=3.86\times10^{5}$ M$_{\rm \odot}/h$ and the total number of
  particles is equal to $2\times384^3$.
\end{itemize}

We ran all the simulations using the {\textit{raijin}}, {\textit
  {vayu}} and {\textit{xe}} clusters at the National Computational
Infrastructure (NCI) National Facility\footnote{http://nf.nci.org.au}
at the Australian National University (ANU). Simulations with box size
equal to 24 Mpc/$h$ typically took from 4.5 to 6.5 computational days
(depending on the configuration) to reach $z=2$, running on 128
CPUs. The simulation with the highest resolution (\textit{Ch12\tu
  eA\tu sW}) reached $z=2$ in more than 2.5 computational months,
running on 160 CPUs. For the post-processing we also used the {\textit
  {edward}} High Performance Computing (HPC) cluster at the University
of Melbourne\footnote{https://its.unimelb.edu.au/research/hpc/edward}.

\section{Cosmic star formation rate density}
\label{cosmsfr}

\begin{figure}
  \centering
  \includegraphics[width=8.6cm]{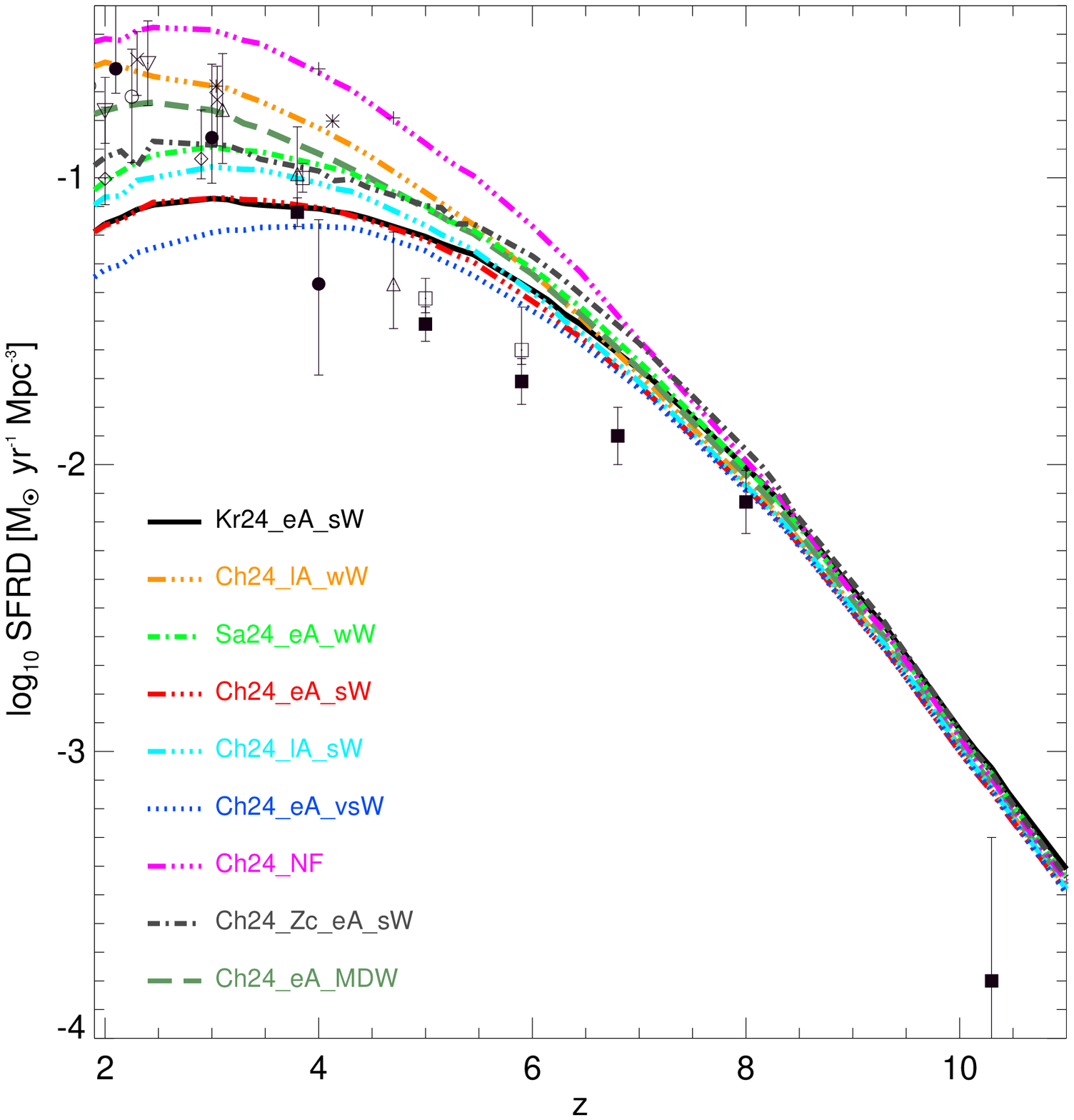}
  \caption{Evolution of the cosmic star formation rate density for all
    the simulations of Table \ref{tab:sim_runs} with box size equal to
    24 Mpc/$h$. The black symbols and error bars refer to different
    observational estimates: \citet{cucciati12} - filled circles;
    \citet{steidel1999} - asterisks; \citet{ouchi04} - plus signs;
    \citet{perez-gonzalez05} - inverted open triangles;
    \citet{schiminovich05} - open diamonds; \citet{bouwens09} - open
    squares; \citet{reddysteidel09} - crosses; \citet{rodighiero10} -
    open circles; \citet{vanderburg10} - upright open triangles;
    \citet{bouwens2012} - filled squares.}
\label{fig_CSFRD_24}
\end{figure}

The evolution of the cosmic star formation rate density is commonly
used to test theoretical models, since it represents a fundamental
constraint on the growth of stellar mass in galaxies over time. In
Figure \ref{fig_CSFRD_24} we plot the cosmic star formation rate
densities (CSFRDs) for all runs with box size $L=24$ Mpc/$h$. Our aim
is to compare the effect of different feedback configurations, choice
of IMF and metal cooling. In Appendix B we show the resolution and box
size tests using simulations with $L=18$ Mpc/$h$ and $L=12$
Mpc/$h$. These tests indicate that the CSFRD numerically converges
below redshift $z\sim4.5$. Overall, our simulations are qualitatively
in agreement with the overplotted observational data. However, we
stress that a direct quantitative comparison should not be made, since
observed CSFRDs are derived by integrating luminosity functions down
to a magnitude limit which depends on redshift and selection criteria,
and is different for different observations.

Since the total integrated amount of gas converted into ``stars'' is
the same for different IMFs, the choice of IMF plays a minor role in
the resulting CSFRD evolution. For example the \textit{Kr24\tu eA\tu
  sW} (black solid line) and the \textit{Ch24\tu eA\tu sW} (red triple
dot-dashed line) simulations, which have exactly the same
configuration aside from the IMF (Kroupa and Chabrier, respectively),
are in good agreement for $z>2$.

All simulations show a similar CSFRD above redshift $z\sim8$, but at
lower redshift start to differ for different feedback
implementations\footnote{Note that at $z=8$ the simulated cosmic star
  formation rate density has not numerically converged. According to
  results presented in Appendix B, the CSFRDs in Figure
  \ref{fig_CSFRD_24} are underestimated.}. Specifically, stronger
feedback results in lower star formation rate density. As expected,
the no-feedback simulation \textit{Ch24\tu NF} (magenta triple
dot-dashed line) shows the highest \mbox{CSFRD}: in this case there is
no effective mechanism able to quench the star formation, and, because
of the ``overcooling'' of gas, too many stars are formed.

Figure \ref{fig_CSFRD_24} illustrates a general trend that at higher
redshift the importance of supernova driven winds increases with
respect to AGN feedback. Galactic winds start to be effective at
$z\le7$. This is visible when the \textit{Sa24\tu eA\tu wW} (early AGN
$+$ weak Winds, light green dot-dashed line) and the \textit{Ch24\tu
  eA\tu sW} (early AGN $+$ strong Winds, red triple dot-dashed line)
runs are compared. In these cases the main difference is related to
the strength of the winds. On the other hand, the AGN feedback is
particularly effective at $z\le5$. This can be seen by comparing the
\textit{Ch24\tu lA\tu sW} run (cyan triple dot-dashed line) with the
\textit{Ch24\tu eA\tu sW} run (red triple dot-dashed line), since the
only difference is in the effectiveness of the AGN feedback. Moreover,
the \textit{Ch24\tu lA\tu sW} run (cyan triple dot-dashed line) falls
below the \textit{Sa24\tu eA\tu wW} (light green dot-dashed line),
suggesting that at high redshift galactic winds regulate AGN
feedback. Finally, the \textit{Ch24\tu eA\tu vsW} (blue dotted line)
and the \textit{Ch24\tu lA\tu wW} (orange triple dot-dashed line) runs
show, respectively, the lowest and the highest CSFRD among runs that
include feedback.

The effect of metal cooling on the CSFRD can be evaluated by comparing
the \textit{Ch24\tu Zc\tu eA\tu sW} run (dot-dashed dark grey line)
with the \textit{Ch24\tu eA\tu sW} run (red triple dot-dashed
line). When metal cooling is included, the star formation rate density
increases at all redshifts and up to a factor of $\sim2$ at
$z=3$. This is due to the fact that in this case the gas can cool more
efficiently via metal-line cooling and forms more stars with respect
to gas with primordial composition.

The CSFRD for the run with momentum-driven galactic winds
\textit{Ch24\tu eA\tu MDW} (dark green dashed line) is in qualitative
agreement with the run with late AGN feedback and weak energy-driven
winds (\textit{Ch24\tu lA\tu wW}, orange triple dot-dashed line). As
we will discuss in the next sections, momentum-driven winds are less
efficient than energy-driven winds in the most massive halos. As a
result, the CSFRD of the \textit{Ch24\tu eA\tu MDW} run at $z\apprle6$
is higher than all the runs including strong energy-driven winds.

\section{Star formation rate functions}
\label{sfrf}

\subsection{Observational data}
\label{obs_sfrf}

In this paper we study the evolution of the star formation rate
function (SFRF) of high redshift galaxies. We compare our simulations
with the observational results of \citet{smit12}, in which the authors
investigated the SFRF of galaxies in the redshift range $z\sim
4-7$. \citet{smit12} adopted two different methods (stepwise and
analytical) to convert UV Luminosity Functions (LFs) into SFR
functions. The two methods are consistent with each other and the
results are well described by a set of Schechter functions
\citep{schechter1976}:
\begin{eqnarray}
  \phi({\rm SFR})\ {\rm d}\,{\rm SFR} & = & \phi^{\star}_{\rm SFR}\left(\frac{{\rm SFR}}{{\rm
        SFR}^{\star}}\right)^{\alpha_{\rm SFR}} \nonumber \\
  & \times & \exp\left(-\frac{{\rm SFR}}{{\rm
        SFR}^{\star}}\right)\frac{{\rm d}\,{\rm SFR}}{{\rm
      SFR}^{\star}}.
\end{eqnarray}
The analytical Schechter parameters ($\phi^{\star}_{\rm SFR}$,
SFR$^{\star}$ and $\alpha_{\rm SFR}$) of \citet{smit12} are shown in
Table \ref{tab_schechtpar}. We also report the stepwise determinations
of the star formation rate function at $z\sim4-7$ in Appendix A (Table
\ref{tab_stepsfrf}).

We note that \citet{salim2012} showed that SFR functions cannot be
adequately described by standard Schechter functions, but are better
described by ``extended'' Schechter functions (where the exponential
part of the standard Schechter function becomes the S\'ersic function)
or Saunders functions \citep{saunders1990}. However, this does not
affect the conclusions from this work.

\begin{figure*}
\centering 
\includegraphics[scale=0.85]{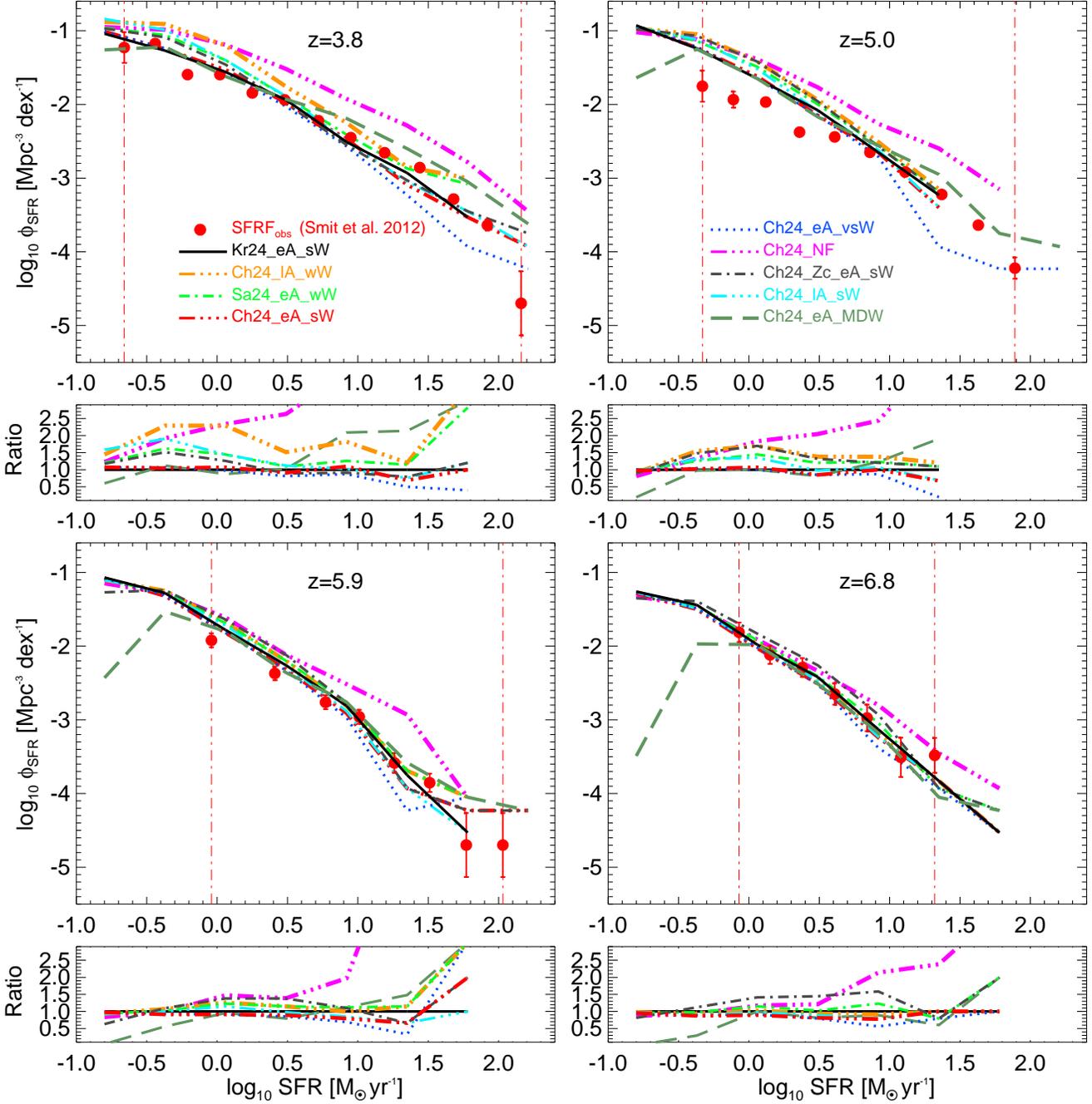}
\caption{Star formation rate functions at $z\sim4-7$ for all the
  simulations of Table \ref{tab:sim_runs} with box size equal to 24
  Mpc/$h$. The red filled circles with error bars and the vertical red
  dot-dashed lines represent the stepwise determinations of the SFR
  function and the observational limits of \citet{smit12},
  respectively. At each redshift, a panel showing ratios between the
  different simulations and the \textit{Kr24\tu eA\tu sW} run (black
  solid line) is included. Line-styles of different simulations are as
  in Figure \ref{fig_CSFRD_24}.}
\label{fig_SFRF}
\end{figure*}

\subsection{Simulated SFRFs}
\label{sim_sfrf}

In Figure \ref{fig_SFRF} we show the star formation rate functions at
redshift $z\sim4-7$ for all runs with box size $L=24$ Mpc/$h$, and
compare these with the results of \citet[][red filled circles with
error bars]{smit12}. The vertical red dot-dashed lines are the
observational limits in the range of SFR. At each redshift, a panel
showing ratios between the different simulations and the
\textit{Kr24\tu eA\tu sW} run (black solid line) is included. In
Appendix B we also perform box size and resolution tests using
simulations with $L=18$ Mpc/$h$ and $L=12$ Mpc/$h$. These tests show
that, while the simulated cosmic star formation rate density converges
only at $z\sim4.5$, the SFR in collapsed structures with mass ${\rm
  M}\ge10^{9.6}$ M$_{\rm \odot}/h$ is convergent out to $z\sim7$.
Since this mass range corresponds to galaxies above the lower limit of
SFR in \citet{smit12}, our results are robust at all redshifts
considered. In this section we first discuss the simulations at
different redshifts, before concentrating on individual properties of
the star formation and feedback schemes.

\begin{table}
\centering
\begin{tabular}{cccccccc}
  \hline & $\langle z\rangle$ & $\phi^{\star}_{\rm SFR}\
  \left(10^{-3}\ {\rm Mpc}^{-3}\right)$ & $\log\frac{{\rm SFR}^{\star}}{{\rm
      M}_{\rm \odot}\ {\rm yr}^{-1}}$ & $\alpha_{\rm SFR}$
  \\ \hline 
  & 3.8 & 1.07$\pm$0.17 & 1.54$\pm$0.10 & -1.60$\pm$0.07  \\
  & 5.0 & 0.76$\pm$0.23 & 1.36$\pm$0.12 &  -1.50$\pm$0.12 \\
  & 5.9 & 1.08$\pm$0.39 & 1.07$\pm$0.17 & -1.57$\pm$0.22 \\
  & 6.8 & 0.64$\pm$0.56 & 1.00$\pm$0.30 & -1.96$\pm$0.35 \\
  \hline \\
\end{tabular}
\caption{Parameters of the Schechter-like approximation
  ($\phi^{\star}_{\rm SFR}$, SFR$^{\star}$ and $\alpha_{\rm SFR}$) to represent the star
  formation rate functions of galaxies at $z\sim 4-7$ from \citet{smit12}.}
\label{tab_schechtpar}
\end{table}

At redshift $z=6.8$ (bottom right panel of Figure \ref{fig_SFRF}) the
no-feedback run (magenta triple dot-dashed line), overproduces the
number of systems in the high SFR tail ($\log\,\frac{{\rm SFR}}{{\rm
    M}_{\odot}\ {\rm yr}^{-1}}\apprge0.8$) with respect to all the
other simulations, due to the overcooling of gas. In this SFR range,
the no-feedback run is marginally consistent with the
observations. All the other runs are consistent with each other and
with the observations, regardless the configuration used for strength
of feedback and choice of IMF. This means that at $z\sim7$ our
simulations are not able to discriminate different schemes of
feedback. The simulation with metal cooling included (\textit{Ch24\tu
  Zc\tu eA\tu sW} - dark grey dot-dashed line) does show an increase
of the SFRF for systems with\footnote{Here and below in the text the
  lower SFR limit corresponds to the lower limit of the observational
  data at the redshift considered (see Table \ref{tab_stepsfrf} in
  Appendix A).} $-0.07<\log\,\frac{{\rm SFR}}{{\rm M}_{\odot}\ {\rm
    yr}^{-1}}\apprle1.0$. As stated in Section \ref{cosmsfr}, this is
due to the fact that when the metals are included in the cooling
function, the gas can cool more efficiently and produce more stars
than gas of primordial composition. As a result, there are more halos
inside the observational window and an increased value of the
SFRF. Inside the observational limits, the simulation with
momentum-driven winds (\textit{Ch24\tu eA\tu MDW} - dark green dashed
line) is in agreement with all the energy-driven wind runs. However,
the momentum-driven scaling of the wind mass loading factor results in
a great suppression of the SFRF in low mass halos. We will discuss the
difference between costant winds and momentum-driven winds in detail
in Sub-section \ref{envsmomw}.

\begin{figure*}
\centering 
\includegraphics[scale=0.85]{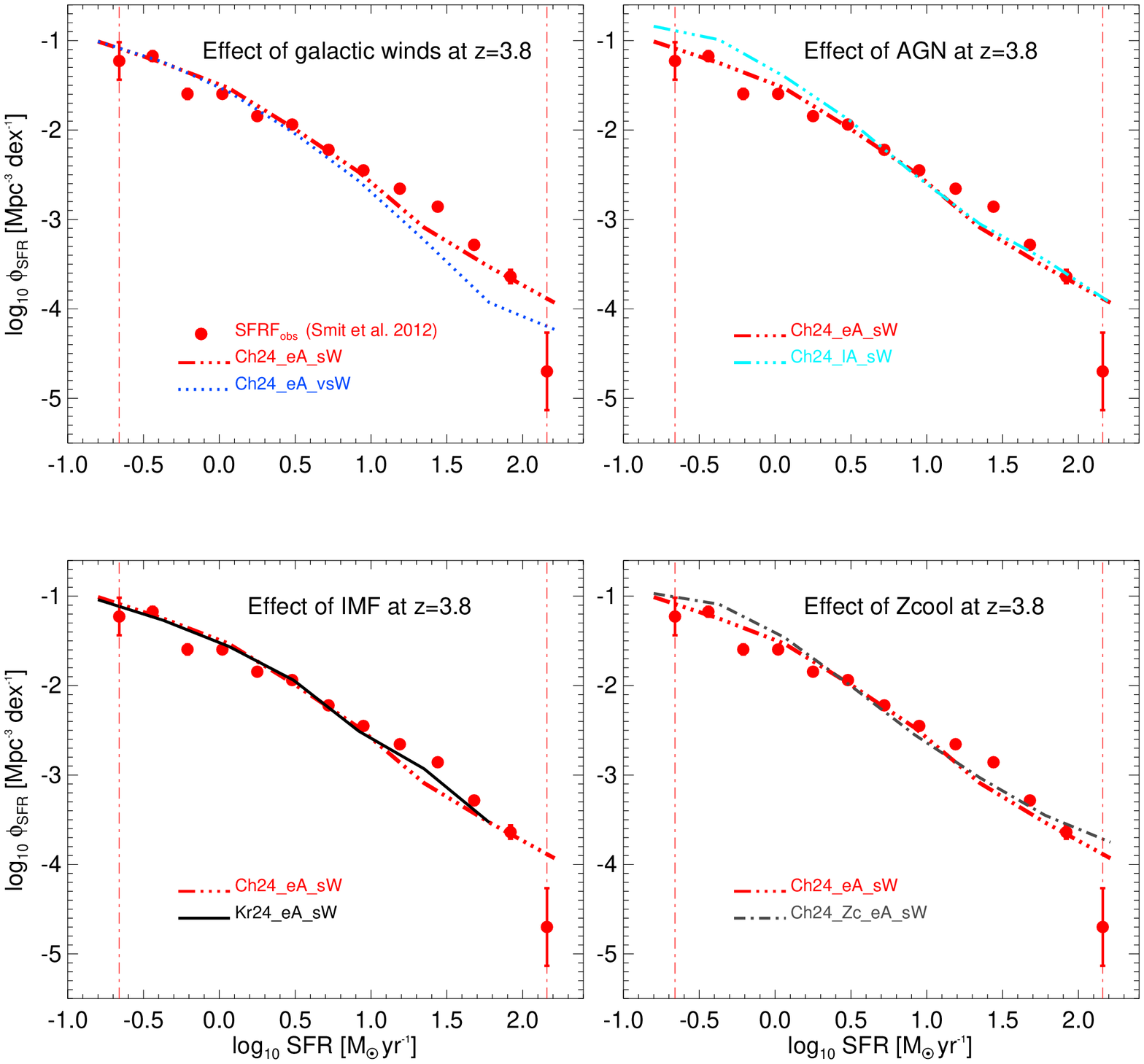}
\caption{Star formation rate functions at redshift
  $z=3.8$. {\textit{Top left panel}}: effect of galactic winds.
  {\textit{Top right panel}}: effect of AGN feedback. {\textit{Bottom
      left panel}}: effect of IMF. {\textit{Bottom right panel}}:
  effect of metal cooling. In each panel, the red filled circles with
  error bars and the vertical red dot-dashed lines represent the
  stepwise determination of the SFR function and the observational
  limits of \citet{smit12}, respectively.}
\label{fig_SFRF_comp}
\end{figure*}

A similar trend is seen at redshift $z=5.9$ (bottom left panel). Aside
from the no-feedback case, we again have good agreement between
observations and simulations, though the simulations slightly
overproduce objects with low star formation rates.

For $z=5.0$ (top right panel) we predict more galaxies with low and
medium SFRs ($\log\,\frac{{\rm SFR}}{{\rm M}_{\odot}\ {\rm
    yr}^{-1}}\apprle0.5$) than observed. The effect of the different
feedback mechanisms starts to become more visible at this
redshift. Simulations with weak feedback (\textit{Ch24\tu lA\tu sW};
\textit{Sa24\tu eA\tu wW}; \textit{Ch24\tu lA\tu wW}) show an excess
of systems with SFR in the range $-0.33<\log\,\frac{{\rm SFR}}{{\rm
    M}_{\odot}\ {\rm yr}^{-1}}\apprle0.8$, with respect to simulations
with strong feedback (\textit{kr24\tu eA\tu sW}; \textit{Ch24\tu eA\tu
  sW}; \textit{Ch24\tu eA\tu vsW}) and the momentum-driven wind run
\textit{Ch24\tu eA\tu MDW}. The simulation with metal cooling
(\textit{Ch24\tu Zc\tu eA\tu sW}) again shows an increase of the SFRF
at $\log\,\frac{{\rm SFR}}{{\rm M}_{\odot}\ {\rm yr}^{-1}}\apprle1.0$,
with respect to the corresponding simulation without metal cooling
(\textit{Ch24\tu eA\tu sW}).

Finally, the behaviour of the different simulations becomes more clear
at $z=3.8$ (top left panel). The no-feedback run overproduces systems
throughout the observational window. At this redshift, it is possible
to distinguish the relative impact of different feedback
mechanisms. In Figure \ref{fig_SFRF_comp} we highlight the influence
of different forms of feedback, metal cooling and IMF on the SFRF at
$z=3.8$. We discuss these different effects in the following
sub-sections.

\subsubsection{Effect of feedback}

To isolate the effects of feedback at $z=3.8$ we only consider those
simulations which have a Chabrier IMF, energy-driven winds and no
metal cooling. Among these simulations, the high SFR tail
($\log\,\frac{{\rm SFR}}{{\rm M}_{\odot}\ {\rm yr}^{-1}}\apprge1.3$)
of the run with late AGN feedback and weak winds (\textit{Ch24\tu
  lA\tu wW}) produces the highest values of the SFRF. The second and
the third highest both have strong winds but late and early AGN
feedback, respectively (\textit{Ch24\tu lA\tu sW} and \textit{Ch24\tu
  eA\tu sW}). The latter two simulations do not show any difference at
high SFRs (even though the implementation of the AGN feedback is
different). However, they have lower SFRF than the \textit{Ch24\tu
  lA\tu wW} model. Finally, the very strong winds case
(\textit{Ch24\tu eA\tu vsW} - blue dotted line) has the lowest value
of the SFRF in the high SFR tail (this feature is already visible at
$z=5.0$). The top left panel of Figure \ref{fig_SFRF_comp} shows the
isolated effect of SN feedback by comparing \textit{Ch24\tu eA\tu sW}
with \textit{Ch24\tu eA\tu vsW}.

The situation is different at low star formation rates
($\log\,\frac{{\rm SFR}}{{\rm M}_{\odot}\ {\rm
    yr}^{-1}}\apprle0.2$). In this range, the \textit{Ch24\tu lA\tu
  wW} run produces more systems with respect to the other three
runs. However, the \textit{Ch24\tu lA\tu sW} run and the
\textit{Ch24\tu eA\tu sW} run are not equal at these low SFRs. The
SFRF of the early AGN simulation is lowered, and agrees well with the
very strong winds run (\textit{Ch24\tu eA\tu vsW}). This suggests that
at $z=3.8$ the AGN feedback in our simulations is important in shaping
the SFRF in the low SFR range.  We discuss in Section \ref{discussion}
how this is related to our black hole seeding scheme. The effect of
AGN feedback can be clearly seen in the top right panel of Figure
\ref{fig_SFRF_comp}.

\begin{figure*}
\centering 
\includegraphics[scale=0.85]{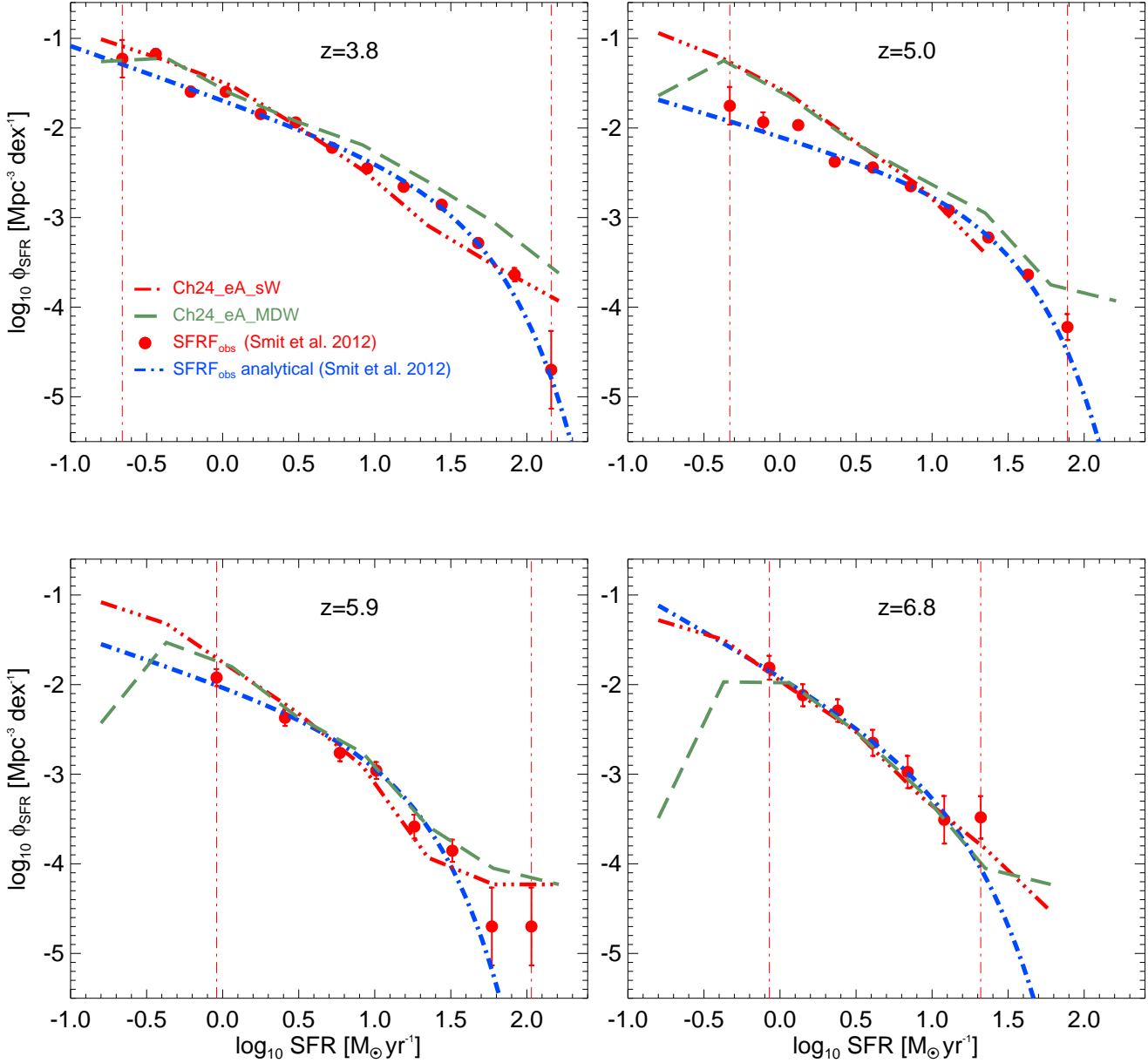}
\caption{Star formation rate functions at $z\sim4-7$: comparison of
  energy- and momentum-driven galactic winds. Red triple dot-dashed
  lines: \textit{Ch24\tu eA\tu sW} run with Chabrier initial mass
  function, early AGN feedback and energy-driven galactic winds of
  (constant) velocity $v_{\rm w}=450$ km/s. Dark green dashed lines:
  \textit{Ch24\tu eA\tu MDW} run with Chabrier initial mass function,
  early AGN feedback and momentum-driven galactic winds ($v_{\rm
    w}=2\times v_{\rm circ}$). The blue dot-dashed lines are the
  observed analytical Schechter-like SFRFs shown in Table
  \ref{tab_schechtpar}. The red filled circles with error bars are the
  stepwise determinations of the SFRF shown in Table
  \ref{tab_stepsfrf} (Appendix A). The vertical red dot-dashed lines
  mark the observational limits. All the observational results are
  from \citet{smit12}.}
\label{fig_SFRF_wind_somp}
\end{figure*}

\subsubsection{Effect of IMF}

At all redshifts, the choice of IMF has only a minor impact on the
SFRF. By comparing the \textit{Ch24\tu eA\tu sW} and the
\textit{Kr24\tu eA\tu sW} runs we see agreement at all SFRs, apart
from a small deviation at the high SFR tail, where the SFRF of the
Chabrier IMF falls below the SFRF of the Kroupa IMF (see bottom left
panel of Figure \ref{fig_SFRF_comp}). The fact that the IMF has a
marginal impact is not surprising, given that it affects mostly metal
production and metal-line cooling is not included in these two
simulations. In principle, changing the IMF should also change the
number of SN produced and, therefore, the corresponding budget of
energy feedback. However, since in our kinetic feedback model we fix
both wind velocity and mass upload rate, the corresponding efficiency
is not related to the number of SN. In fact, this number enters only
as thermal feedback to regulate star formation in the ISM effective
model. Since this feedback channel is quite inefficient, the net
result is that the IMF does not have a sizeable effect on the SFR.

\subsubsection{Effect of metal cooling}

By comparing the \textit{Ch24\tu Zc\tu eA\tu sW} and the
\textit{Ch24\tu eA\tu sW} runs we are able to check the influence of
metal cooling on the simulated SFRFs. At $z\ge5$, metal cooling is
responsible for the increase in the number of objects with
$\log\,\frac{{\rm SFR}}{{\rm M}_{\odot}\ {\rm yr}^{-1}}\apprle1.0$. At
$z=3.8$, we can see that this increment is significant only in the low
star formation rate tail of the distribution (see bottom right panel
of Figure \ref{fig_SFRF_comp}). At this redshift, the effect of metal
cooling is less important than the effect of different feedback
prescriptions.

\subsubsection{Relative importance of galactic winds and AGN feedback}

Since the choice of IMF plays a minor role on the SFRF, we next
compare the \textit{Sa24\tu eA\tu wW} (Salpeter IMF) with the
\textit{Ch24\tu lA\tu wW} and the \textit{Ch24\tu lA\tu sW} runs
(Chabrier IMFs), in the top left panel of Figure \ref{fig_SFRF}. At
high SFRs the \textit{Sa24\tu eA\tu wW} and the \textit{Ch24\tu lA\tu
  wW} are in agreement. These simulations have the same wind strength
but different AGN implementations. The \textit{Ch24\tu lA\tu sW} has
lower SFRF than the two weak wind models. At low SFRs, the SFRF values
are ranked according to: ${\rm SFRF}_{\textit{Ch24\_lA\_wW}}>{\rm
  SFRF}_{\textit{Ch24\_lA\_sW}}>{\rm
  SFRF}_{\textit{Sa24\_eA\_wW}}$. This suggests that, in our
simulations at high redshift, galactic winds shape the SFRF over the
whole range of star formation rates and their effect is more important
than AGN feedback for halos with mass M $\apprle10^{12}$ M$_{\rm
  \odot}$/$h$ (i.e. the mass of the most massive halo at $z=3.8$). As
a result of our black hole seeding scheme, the effect of AGN feedback
is most visible at low SFRs (see the discussion in Section
\ref{discussion}).

\subsubsection{Constant vs. momentum-driven galactic winds}
\label{envsmomw}

In Figure \ref{fig_SFRF_wind_somp} we compare the evolution of the
SFRF for the \textit{Ch24\tu eA\tu sW} run (Chabrier IMF, early AGN
feedback and energy-driven galactic winds of constant velocity $v_{\rm
  w}=450$ km/s - red triple dot-dashed lines) and the \textit{Ch24\tu
  eA\tu MDW} run (Chabrier IMF, early AGN feedback and momentum-driven
galactic winds - dark green dashed lines). In the figure, besides the
stepwise determinations of the SFRF from \citet[][red filled circles
with error bars]{smit12}, we show the analytical Schechter-like SFRFs
from the same work (blue dot-dashed lines; the parameters of these
analytical functions are presented in Table \ref{tab_schechtpar}). The
vertical red dot-dashed lines mark the observational limits.

At $z=6.8$ the two simulations are in good agreement inside the
observational window. At $z=5.9$ a slight excess of systems at
$\log\,\frac{{\rm SFR}}{{\rm M}_{\odot}\ {\rm yr}^{-1}}\apprge1.0$ is
visible for the momentum-driven wind run. Moreover, as we pointed out
in Section \ref{sim_sfrf}, the momentum-driven scaling of the wind
mass loading factor results in a great suppression of the SFRF in low
mass halos (although outside the observational limits). At lower
redshift, as the mass function moves towards larger masses, this
suppression becomes less pronounced and almost disappears at $z=3.8$.

At $z=5.0$ (3.8) the \textit{Ch24\tu eA\tu sW} and the \textit{Ch24\tu
  eA\tu MDW} runs produce the same SFRF for $\log\,\frac{{\rm
    SFR}}{{\rm M}_{\odot}\ {\rm yr}^{-1}}\apprle0.8$
($\log\,\frac{{\rm SFR}}{{\rm M}_{\odot}\ {\rm
    yr}^{-1}}\apprle0.5$). However, in massive halos momentum-driven
winds are less efficient than energy-driven winds in quenching the
star formation rate. Consequently, in the high end of the distribution
${\rm SFRF}_{\textit{Ch24\_eA\_MDW}}>{\rm
  SFRF}_{\textit{Ch24\_eA\_sW}}$. It is straightforward to understand
this trend for the \textit{Ch24\tu eA\tu MDW} simulation. In small
halos the velocity of the winds is low, but their efficiency is large
($\eta\propto v_{\rm w}^{-1}$). Therefore, even low velocity winds are
efficient in stopping the ongoing star formation rate. On the other
hand, in massive halos winds have velocities $v_{\rm w}\gg 450$ km/s
and small loading factors ($\eta\ll2$). For this reason, only a few
wind particles are created and these are not sufficient to effectively
suppress the formation of stars (even if they can easily escape from
galaxies and reach the IGM). Overall, the momentum-driven wind
simulation is consistent with the observations of \citet{smit12}.

\begin{figure}
  \centering
  \includegraphics[width=8.6cm]{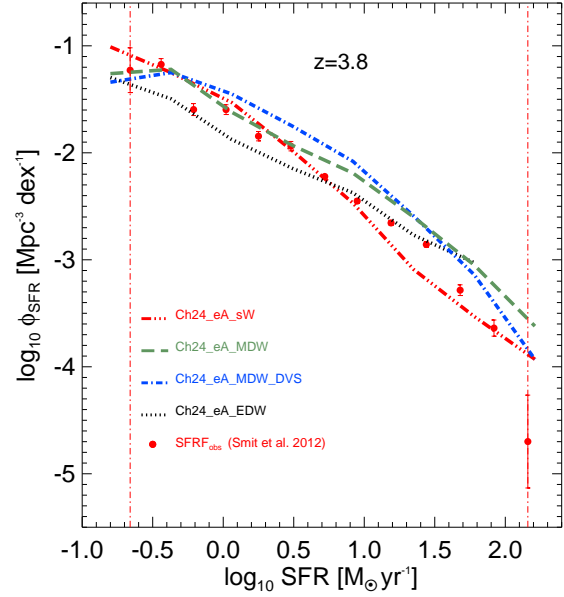}
  \caption{Star formation rate function at $z=3.8$: comparison of
    different choices of parameters for the energy- and
    momentum-driven galactic winds. All the simulations shown have a
    Chabrier initial mass function and early AGN feedback. Red triple
    dot-dashed line: \textit{Ch24\tu eA\tu sW} run with energy-driven
    galactic winds of constant velocity $v_{\rm w}=450$ km/s and
    loading factor $\eta=2$. Dark green dashed line: \textit{Ch24\tu
      eA\tu MDW} run with momentum-driven winds ($v_{\rm w}=2\times
    v_{\rm circ}$ and $\eta=2\times\frac{450\,\,{\rm km/s}}{v_{\rm
        w}}$). Blue dot-dashed line: \textit{Ch24\tu eA\tu MDW\tu DVS}
    run with momentum-driven winds and different velocity scaling
    ($v_{\rm w}=v_{\rm circ}$ and $\eta=2\times\frac{450\,\,{\rm
        km/s}}{v_{\rm w}}$). Black dotted line: \textit{Ch24\tu eA\tu
      EDW} run with energy-driven winds ($v_{\rm w}=2\times v_{\rm
      circ}$ and $\eta=2\times\left(\frac{450\,\,{\rm km/s}}{v_{\rm
          w}}\right)^2$). The last two simulations have been run only
    as a comparison and are not part of the set discussed throughout
    this paper. The red filled circles with error bars and the
    vertical red dot-dashed lines represent, respectively, the
    stepwise determination of the SFR function and the observational
    limits of \citet{smit12} at $z=3.8$.}
\label{fig_EDWMDW}
\end{figure}

For further comparison, we performed several tests by changing the
velocity of the winds ($v_{\rm w}=v_{\rm circ}$ instead of $v_{\rm
  w}=2\times v_{\rm circ}$: run \textit{Ch24\tu eA\tu MDW\tu DVS},
where \textit{DVS} stands for ``Different Velocity Scaling''), and
also using the energy-driven scaling of the wind mass loading factor
adopted by \citet{puchespring13}: $\eta\propto v_{\rm w}^{-2}$ (run
\textit{Ch24\tu eA\tu EDW})\footnote{Note that the \textit{Ch24\tu
    eA\tu MDW\tu DVS} and \textit{Ch24\tu eA\tu EDW} simulations have
  been run only for the sake of these tests and are not part of the
  set discussed throughout the paper.}. The result of these tests is
shown in Figure \ref{fig_EDWMDW}. Our conclusion is that in order to
reproduce the SFRFs at high redshift, a non-aggressive variable wind
scaling is needed, otherwise the number of objects with low SFRs is
greatly suppressed, while at the same time winds are not effective in
the most massive systems. As a consequence, the SFRFs from the
\textit{Ch24\tu eA\tu MDW} run are in a qualitative agreement with the
constant wind models used in this work.

\begin{figure*}
\centering 
\includegraphics[scale=0.85]{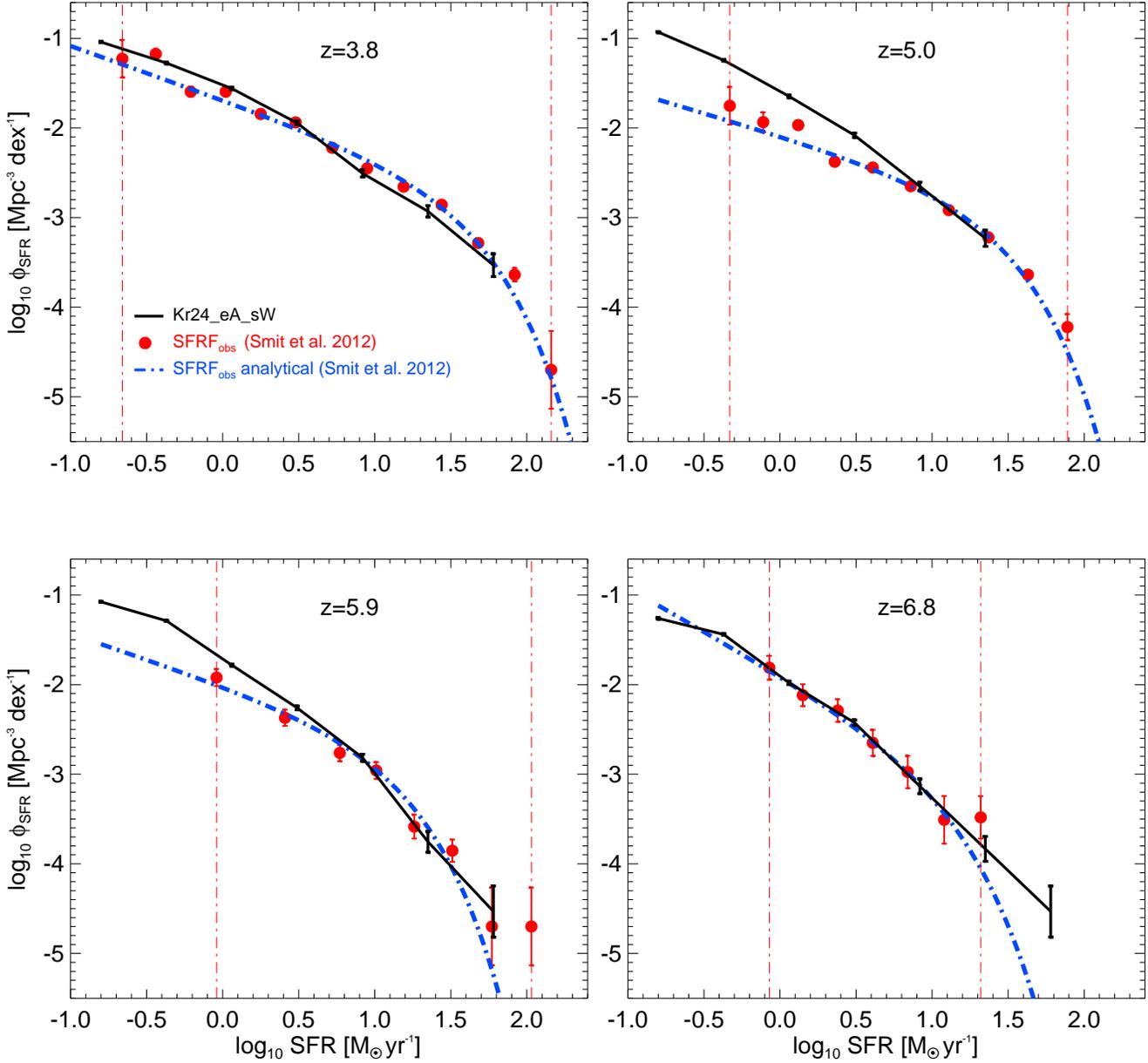}
\caption{Star formation rate functions at $z\sim4-7$ for our best
  model with Kroupa initial mass function, early AGN feedback and
  strong energy-driven galactic winds of velocity $v_{\rm w}=450$ km/s
  (black solid lines). The black error bars are the Poissonian
  uncertainties of the simulated SFRFs. The blue dot-dashed lines are
  the observed analytical Schechter-like SFRFs shown in Table
  \ref{tab_schechtpar}.  The red filled circles with error bars are
  the stepwise determinations of the SFRF shown in Table
  \ref{tab_stepsfrf} (Appendix A). The vertical red dot-dashed lines
  mark the observational limits. All the observational results are
  from \citet{smit12}.}
\label{fig_SFRF_kr24}
\end{figure*}

\subsubsection{Best model}

In Figure \ref{fig_SFRF_kr24} we show the star formation rate
functions at redshift $z\sim4-7$ for our best model: \textit{Kr24\tu
  eA\tu sW} (Kroupa IMF, early AGN feedback and strong energy-driven
galactic winds with $v_{\rm w}=450$ km/s). This simulation provides
the best description of the observations among the models considered
in our analysis, especially at $z=3.8$. However, like all the other
runs that include feedback, at $z=5.0$ it overproduces objects with
$\log\,\frac{{\rm SFR}}{{\rm M}_{\odot}\ {\rm
    yr}^{-1}}\apprle0.5$. This either suggests a bias in the
observations, or that the calibration of galactic winds and AGN
feedback must be varied in order to reproduce the observational data
over the redshift range $4\le z\le7$.

In Figure \ref{fig_SFRF_kr24}, the blue dot-dashed lines and the red
filled circles with error bars are, respectively, the analytical
Schechter-like SFRFs and the stepwise determinations of the SFRF of
\citet{smit12}. The vertical red dot-dashed lines mark the
observational limits. We also include the Poissonian uncertainties for
the simulated SFRFs (black error bars), in order to provide an
estimate of the errors from our finite box size. The uncertainties are
larger at high SFRs due to the small number of massive halos in the
box.

\section{Discussion}
\label{discussion}

At the highest redshift considered in this work, we find that our
simulations are not able to distinguish the effect of different
feedback prescriptions. In fact, at $z\sim7$ all runs with feedback
included reproduce the observational data. On the other hand, the case
with no feedback fails to fit the observations. An important
conclusion is therefore that feedback effects start to be important
from very high redshift. Moving to lower redshift (and especially at
$z=3.8$), different feedback configurations show different trends,
implying that the SFRF can be used to probe the physics of star
formation and feedback at high $z$. As discussed in Section
\ref{cosmsfr}, supernova driven galactic winds start to be effective
at $z\le7$.
% Therefore, they shape the SFRF and regulate AGN feedback.

In the top left panel of Figure \ref{fig_SFRF_comp} the
\textit{Ch24\tu eA\tu sW} run and the \textit{Ch24\tu eA\tu vsW} run
are compared at $z=3.8$. They have exactly the same configuration
except for the wind velocity (450 km/s and 550 km/s,
respectively). The two SFRFs are in agreement for $\log\,\frac{{\rm
    SFR}}{{\rm M}_{\odot}\ {\rm yr}^{-1}}\apprle1.0$, while in the
high SFR tail ${\rm SFRF}_{\textit{Ch24\_eA\_sW}}>{\rm
  SFRF}_{\textit{Ch24\_eA\_vsW}}$. This indicates that, for halos with
$\log\,\frac{{\rm SFR}}{{\rm M}_{\odot}\ {\rm yr}^{-1}}\apprle1.0$,
the effect of the two wind configurations is the same: both
efficiently remove gas particles from the central regions and kick
them out of the collapsed structures. At higher SFRs/masses, weaker
winds become less effective in expelling gas and wind particles remain
trapped within halos. We stress that in our energy-driven scheme, even
if we consider different wind velocities, the efficiency of the winds
is fixed to $\eta=\dot{\rm M}_{\rm w}/\dot{\rm M}_{\star}=2$ (see
Eq. \ref{eq_windload}). As a consequence, for different configurations
the wind mass loading ($\dot{\rm M}_{\rm w}$) is the same at a given
star formation rate ($\dot{\rm M}_{\star}$). Since we also decouple
wind particles from the hydrodynamics for a certain period of time,
ejecta at higher velocities are more effective in removing gas from
halos. This in turn explains why ${\rm
  SFRF}_{\textit{Ch24\_eA\_sW}}>{\rm SFRF}_{\textit{Ch24\_eA\_vsW}}$
in the high SFR tail of the distribution.

We compared the constant galactic wind models with a momentum-driven
variable wind scheme. In this model, the velocity of the wind depends
on the circular velocity of the halo $v_{\rm w}= 2\times v_{\rm
  circ}$. The mass loading factor $\eta = 2\times\frac{450\,\,{\rm
    km/s}}{v_{\rm w}}$ ($\eta=2$ if the wind velocity is equal to our
reference constant ``strong'' wind model $v_{\rm w}=450$
km/s). Overall, the momentum-driven wind simulation is in agreement
with the constant wind models and consistent with the observations of
\citet{smit12}.

While at redshift $z\ge3.8$ galactic winds are already in place and
dominate the feedback mechanisms, AGN feedback is not yet very
efficient. In our AGN model, we seed all the halos above a given mass
threshold (M$_{\rm th}$) with a central SMBH of mass M$_{\rm seed}$,
provided they contain a minimum mass fraction in stars $f_{\star}$
(see the end of Sub-section \ref{agn_feedb}). These SMBHs can then
grow via gas accretion and through mergers with other SMBHs. Our
simulations explored two regimes for the AGN feedback, with varied
M$_{\rm th}$, $f_{\star}$, M$_{\rm seed}$ and the maximum accretion
radius $R_{\rm ac}$. In the ``early AGN'' configuration we reduced
M$_{\rm th}$, $f_{\star}$ and M$_{\rm seed}$ and increased $R_{\rm
  ac}$, with respect to the ``late AGN'' configuration. However, the
radiative efficiency ($\epsilon_{\rm r}$) and the feedback efficiency
($\epsilon_{\rm f}$) are the same in the two regimes. Decreasing the
threshold mass for seeding a SMBH increases the effect of AGN feedback
on halos with low SFRs by construction, since we allow the presence of
a black hole in lower mass halos (compare the \textit{Ch24\tu eA\tu
  sW} run and the \textit{Ch24\tu lA\tu sW} run in the upper right
panel of Figure \ref{fig_SFRF_comp}). The case with early AGN leads
asymptotically to the Magorrian relation \citep{magorrian1998} at low
redshift, but imposes high black hole$/$halo mass ratios in small
galaxies at early times. As a result, ${\rm
  SFRF}_{\textit{Ch24\_eA\_sW}}<{\rm SFRF}_{\textit{Ch24\_lA\_sW}}$ in
the low SFR tail of the distribution. On the other hand, there is no
difference between the \textit{Ch24\tu eA\tu sW} run and the
\textit{Ch24\tu lA\tu sW} run at high SFRs. At $z=3.8$, halos with
$\log\,\frac{{\rm SFR}}{{\rm M}_{\odot}\ {\rm yr}^{-1}}\ge1.0$ have
stellar masses $10^{10}$ M$_{\rm \odot}$/$h\le$ M$_{\star}\le10^{11}$
M$_{\rm \odot}$/$h$. For these halos, the central SMBHs have grown to
masses $5\times10^{7}$ M$_{\rm \odot}/h\le$ M$_{\rm
  SMBH}\le5\times10^{8}$ M$_{\rm \odot}$/$h$ and are accreting at a
moderate level. As shown in \citet{Dimatteo}, in these objects the
star formation rate is essentially unaffected by the presence of the
black holes, since the SMBHs have not yet reached a regime of
self-regulating growth. Due to the small box size, at high redshift
our simulations are not able to form cluster-sized objects with total
mass M $\apprge10^{12}$ M$_{\rm \odot}$/$h$, where the black hole
growth is expected to be exponential at early times
\citep{dimatteo08}. In this case, the central SMBHs would grow until
they release sufficient energy to generate outflows and prevent
further star formation.
% \citep[i.e. they would enter the self-regulation regime discussed
% in][]{Dimatteo}.
This explains why we do not see any AGN feedback effect in the high
star formation rate tail of the SFRF. \\

In conclusion, in our simulations the interplay between galactic winds
and AGN feedback suggests that at high redshift SN driven winds are
essential to reproduce the observed SFRFs. According to our scheme,
the effect of AGN feedback in the low end of the luminosity/SFR
functions is sensitive to the seeding of super-massive black holes. We
are extending this work to lower redshift, in order to compare our
results with different observations and other theoretical works. For
example, \citet{dave11} examined the growth of the stellar content of
galaxies at $z=0-3$. They ran four different galactic wind
models. With these simulations, they produced stellar mass and star
formation rate functions to quantify the effects of outflows on the
galactic evolution at low redshift. In their simulations, winds are
responsible for the shape of the faint end slope of the SFR function
at $z=0$ \citep[top left panel of Figure 2 in][]{dave11}. AGN feedback
is not included and, as a result, their simulations overproduce the
number of objects in the high star formation rate tail. We will show
how in our simulations AGN are crucial to reproduce the observed high
end of the luminosity/SFR functions at low redshift ($z\apprle1$), as
pointed out by other authors \citep{puchespring13} and also by
semi-analytic models \citep[e.g.][]{croton06,bower06}.

\section{Conclusions}
\label{concl}

In this paper we have presented a new set of cosmological simulations,
{\textsc{Angus}} ({\textit{AustraliaN {\small{GADGET-3}} early
    Universe Simulations}}), run with the hydrodynamic code
{\small{P-GADGET3(XXL)}}. We have used these to investigate the star
formation rate function of high redshift galaxies ($z\sim4-7$), with
comparison to the observations of \citet{smit12}. In particular, we
have focused on the role of feedback from SN and AGN and studied the
impact of metal cooling and different IMFs. We ran 11 simulations with
various feedback configurations and box sizes ($L=24$, 18 and 12
Mpc/$h$). We used the \citet{springel2003} implementation of SN
energy-driven galactic winds. In particular, we explored three
different configurations: weak, strong and very strong winds of
constant velocity $v_{\rm w}=350$, 450 and 550 km/s,
respectively. Moreover, following \citet{puchespring13}, in one
simulation we adopted variable momentum-driven galactic winds. We also
explored two regimes for the AGN feedback (early and late). The early
AGN scheme imposes high black hole$/$halo mass ratios in small
galaxies at early times. This configuration leads asymptotically to
the Magorrian relation \citep{magorrian1998} at low redshift, but
accentuates the effect of AGN in low SFR/mass halos at high $z$. We
considered three different IMFs \citep{salpeter55,kroupa93,chabrier03}
and the effect of metal cooling (see Section \ref{simout}). We have
performed box size and resolution tests to check the convergence of
the results from our simulations (Appendix B). Overall, these tests
confirm that the SFR in collapsed structures with mass ${\rm M}\ge
10^{9.6}$ M$_{\rm \odot}/h$ is convergent at all the redshifts
considered.

The main results and conclusions of this work can be summarised as
follows:
\begin{itemize}
\item We studied the evolution of the cosmic star formation rate
  density (CSFRD). Galactic winds start to be effective at $z\le7$,
  while the AGN feedback becomes important later on, at $z\le5$. When
  metal cooling is included, the cosmic star formation rate density
  increases at all redshifts and up to a factor of $\sim2$ at
  $z=3$. On the other hand, in our simulations the choice of IMF plays
  a minor role on the CSFRD evolution.
\item We explored the star formation rate functions (SFRFs) of
  galaxies at redshift $z\sim4-7$. At $z=6.8$, our simulations are not
  able to discriminate different feedback prescriptions and all runs
  that include feedback reproduce the observational results of
  \citet{smit12}. However, the no-feedback simulation fails to fit the
  observations, indicating that feedback effects start to be important
  and need to be taken into account from very high redshift. The key
  factor required to reproduce the observed SFRFs at lower redshift,
  is a combination of strong winds and early AGN feedback.
\item In our simulations, supernova driven galactic winds shape the
  SFRF in the whole range of star formation rates. This conclusion is
  not qualitatively dependent on the model for AGN feedback.
\item At all the redshifts considered, the choice of IMF has a minor
  impact on the SFRFs. On the other hand, metal cooling is responsible
  for the increase in the number of objects with low and intermediate
  star formation rates. However, at $z\sim4$ the effect of metal
  cooling is less important than the effect of different feedback
  prescriptions.
\item To reproduce the SFRFs at $z\apprge4$, a non-aggressive variable
  wind scaling is needed, otherwise the amount of objects with low
  SFRs is greatly suppressed and at the same time winds are not
  effective in the most massive systems. As a result, the SFRFs from
  our momentum-driven wind simulation are in a qualitative agreement
  with the constant (energy-driven) wind models used in this work.
\end{itemize}

We are exploring the interplay between galactic winds and AGN feedback
at high redshift in a companion paper \citep{kata13}. In that work we
analyse the stellar mass functions and the star formation
rate$-$stellar mass relations for the sample of galaxies considered in
this work.
% In the future we will extend these studies to lower redshift. In
% fact, it is commonly accepted that at low redshift AGN are crucial
% in order to reproduce the observed high end of the luminosity/SFR
% functions. We will show that our simulations confirm this result.
We are also planning to explore more feedback configurations and in
particular different parameters for the AGN feedback and the new wind
models of \citet{barai13}.

\section*{Acknowledgments}

The authors would like to thank Volker Springel for making available
to us the non-public version of the {\small{GADGET-3}} code. ET would
like to thank Giuseppe Murante, James Bolton, Simon Mutch, Paul Lasky,
Umberto Maio, Bartosz Pindor, Alexandro Saro and Mark Sargent for many
insightful discussions. ET is also thankful for the hospitality
provided by the University of Trieste and the Trieste Astronomical
Observatory, where part of this work was completed. PB and MV are
supported by the FP7 ERC Starting Grant ``cosmoIGM''.  This research
was conducted by the Australian Research Council Centre of Excellence
for All-sky Astrophysics (CAASTRO), through project number
CE110001020. This work was supported by the Flagship Allocation Scheme
of the NCI National Facility at the ANU, by the European Commission’s
Framework Programme 7, through the Marie Curie Initial Training
Network CosmoComp (PITN-GA-2009-238356), by the PRIN-MIUR09 ``Tracing
the growth of structures in the Universe'', and by the PD51 INFN
grant.

\bibliographystyle{mn2e}	
\bibliography{tescari_angus_astroph_v2}

\begin{table}
  \centering
  \begin{tabular}{cccc}
    \hline \\
    & {\large $\log\,\frac{{\rm SFR}}{{\rm M}_{\odot}\ {\rm yr}^{-1}}$} & &
    {\large $\phi_{\rm SFR}\ \left({\rm Mpc}^{-3}\ {\rm
          dex}^{-1}\right)$} \\ \\
    \hline \hline
    & & $z\sim4$ &\\ 
    \hline 
    & -0.66 & & 0.05920$\pm$0.02855  \\
    & -0.44 & & 0.06703$\pm$0.00838 \\
    & -0.21 & & 0.02537$\pm$0.00326  \\
    & 0.02 & & 0.02534$\pm$0.00268  \\
    & 0.25 & & 0.01430$\pm$0.00144  \\
    & 0.48 & & 0.01153$\pm$0.00117 \\
    & 0.72 & & 0.00601$\pm$0.00025  \\
    & 0.95 & & 0.00354$\pm$0.00017  \\
    & 1.19 & & 0.00221$\pm$0.00012  \\
    & 1.44 & & 0.00139$\pm$0.00008  \\
    & 1.68 & & 0.00052$\pm$0.00006  \\
    & 1.92 & & 0.00023$\pm$0.00004  \\
    & 2.16 & & 0.00002$\pm$0.00002  \\
    \hline
    & & $z\sim5$ & \\
    \hline
    & -0.33 & & 0.01766$\pm$0.00858 \\
    & -0.11 & & 0.01161$\pm$0.00294 \\
    & 0.12 & & 0.01076$\pm$0.00121 \\
    & 0.36 & & 0.00420$\pm$0.00046 \\
    & 0.61 & & 0.00362$\pm$0.00040 \\
    & 0.86 & & 0.00224$\pm$0.00014 \\
    & 1.11 & & 0.00121$\pm$0.00008 \\
    & 1.37 & & 0.00060$\pm$0.00006 \\
    & 1.63 & & 0.00023$\pm$0.00002 \\
    & 1.89 & & 0.00006$\pm$0.00002 \\
    \hline
    & & $z\sim6$ & \\ 
    \hline
    & -0.04 & & 0.01197$\pm$0.00262 \\
    & 0.41 & & 0.00426$\pm$0.00089 \\
    & 0.77 & & 0.00173$\pm$0.00037 \\
    & 1.01 & & 0.00110$\pm$0.00024 \\
    & 1.26 & & 0.00026$\pm$0.00008 \\
    & 1.51 & & 0.00014$\pm$0.00004 \\
    & 1.77 & & 0.00002$\pm$0.00002 \\
    & 2.03 & & 0.00002$\pm$0.00002 \\
    \hline
    & & $z\sim7$ & \\
    \hline
    & -0.07 & & 0.01543$\pm$0.00473 \\
    & 0.15 & & 0.00761$\pm$0.00215 \\
    & 0.38 & & 0.00513$\pm$0.00149 \\
    & 0.61 & & 0.00224$\pm$0.00075 \\
    & 0.84 & & 0.00106$\pm$0.00044 \\
    & 1.08 & & 0.00031$\pm$0.00019 \\
    & 1.32 & & 0.00033$\pm$0.00018 \\
    \hline \\
  \end{tabular}
  \caption{Stepwise determinations of the SFR function at $z\sim4-7$ \citep{smit12}.}
  \label{tab_stepsfrf}
\end{table}

\section*{Appendix A: Observed stepwise SFRFs}
\label{appendix_a}

Table \ref{tab_stepsfrf} shows the \citet{smit12} stepwise
determinations of the star formation rate function $\phi_{\rm SFR}$
from dust-corrected UV luminosity functions at $z\sim4-7$.

\section*{Appendix B: Box size and resolution tests}
\label{appendix_b}

In this appendix we perform box sixe and resolution tests, in order to
check the convergence of the results from our simulations. We
underline that in this paper the smaller the box size of a run, the
higher its mass/spatial resolution. However, the box size sets an
upper limit on the mass of the halos that can be formed in the
simulated volume. Therefore, higher resolution means poorer statistics
at the high mass end of the halo mass function.

\begin{figure*}
\centering
\includegraphics[width=8.8cm]{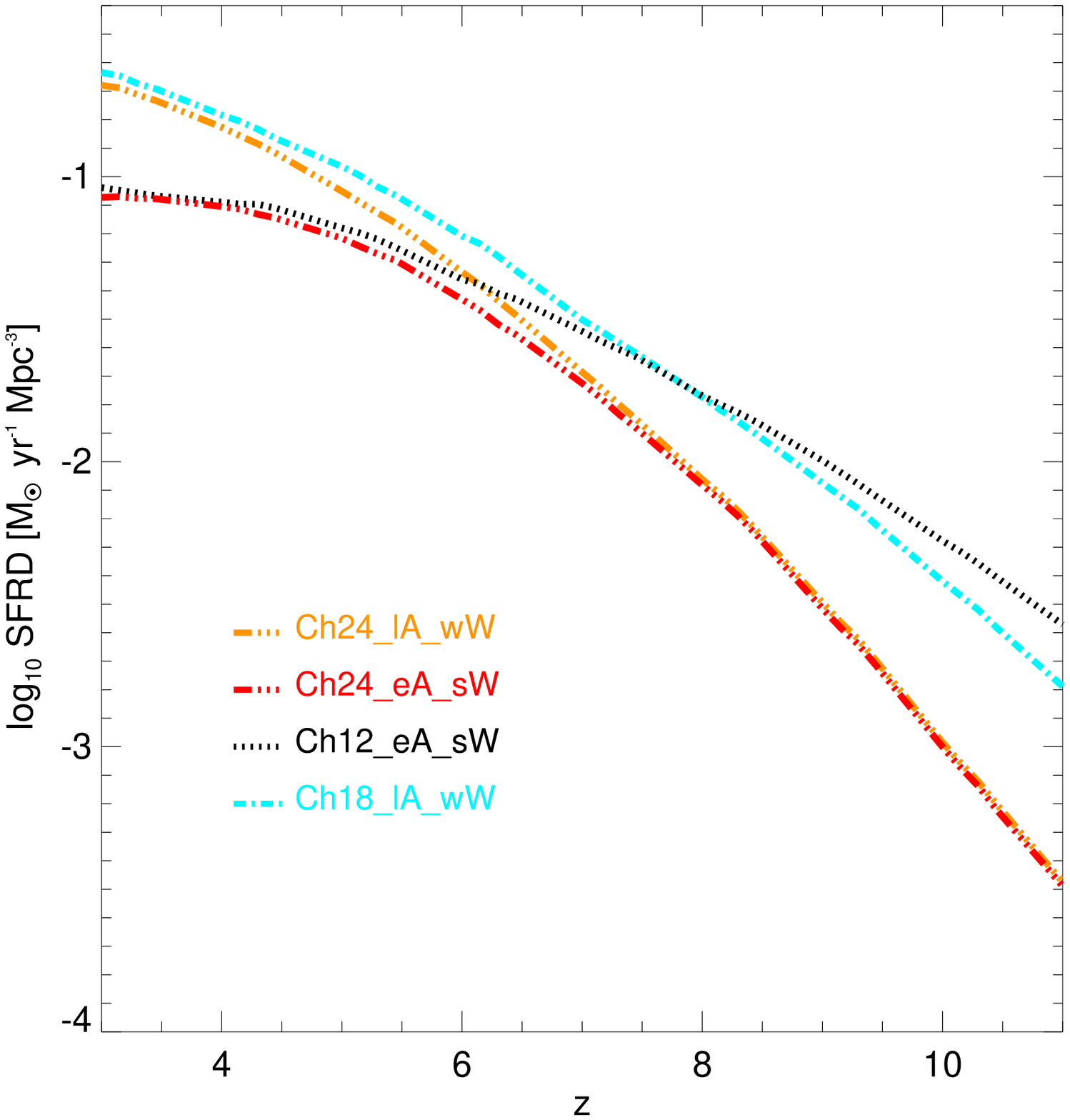} \\
\includegraphics[width=8.8cm]{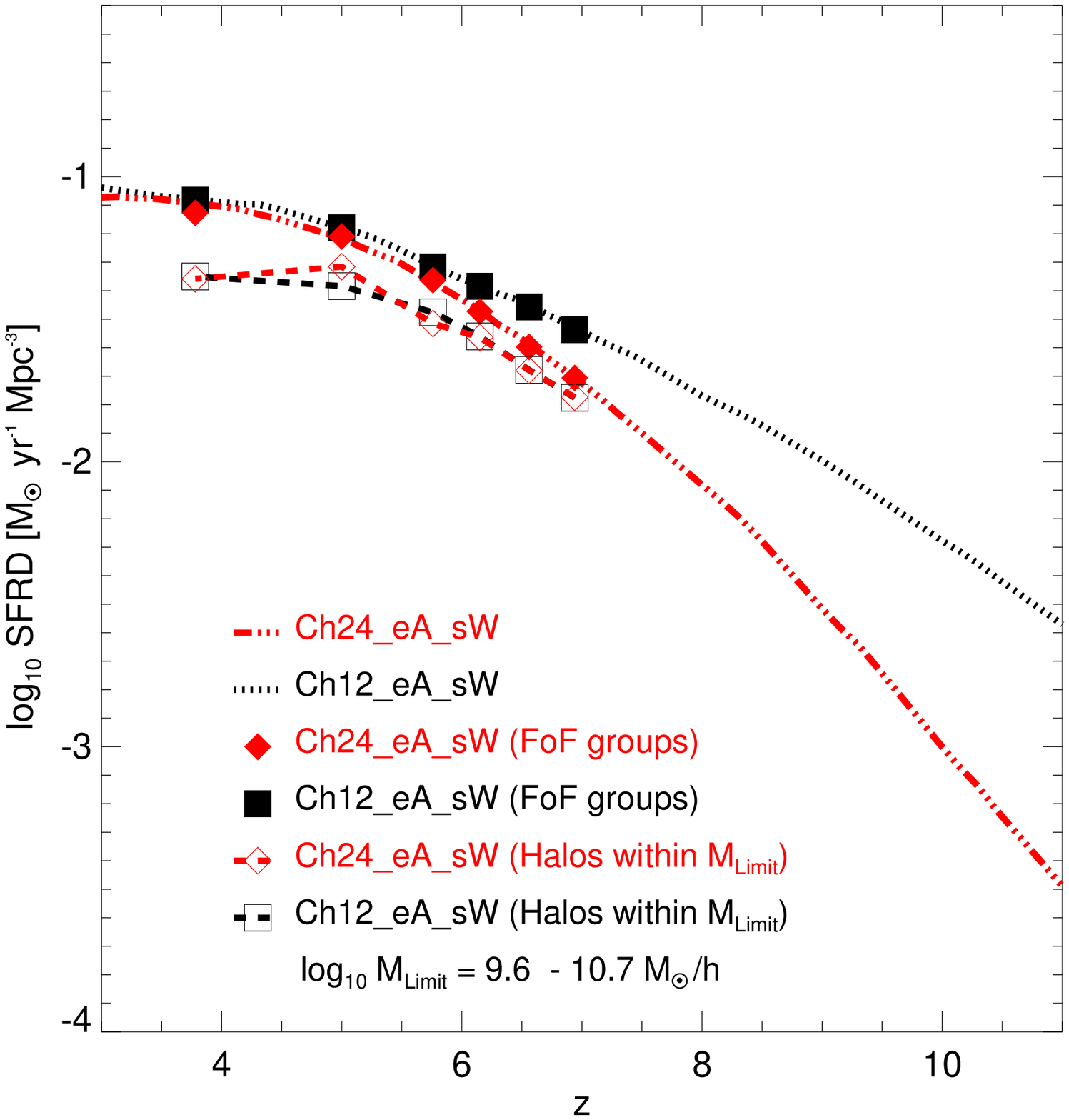}
\includegraphics[width=8.8cm]{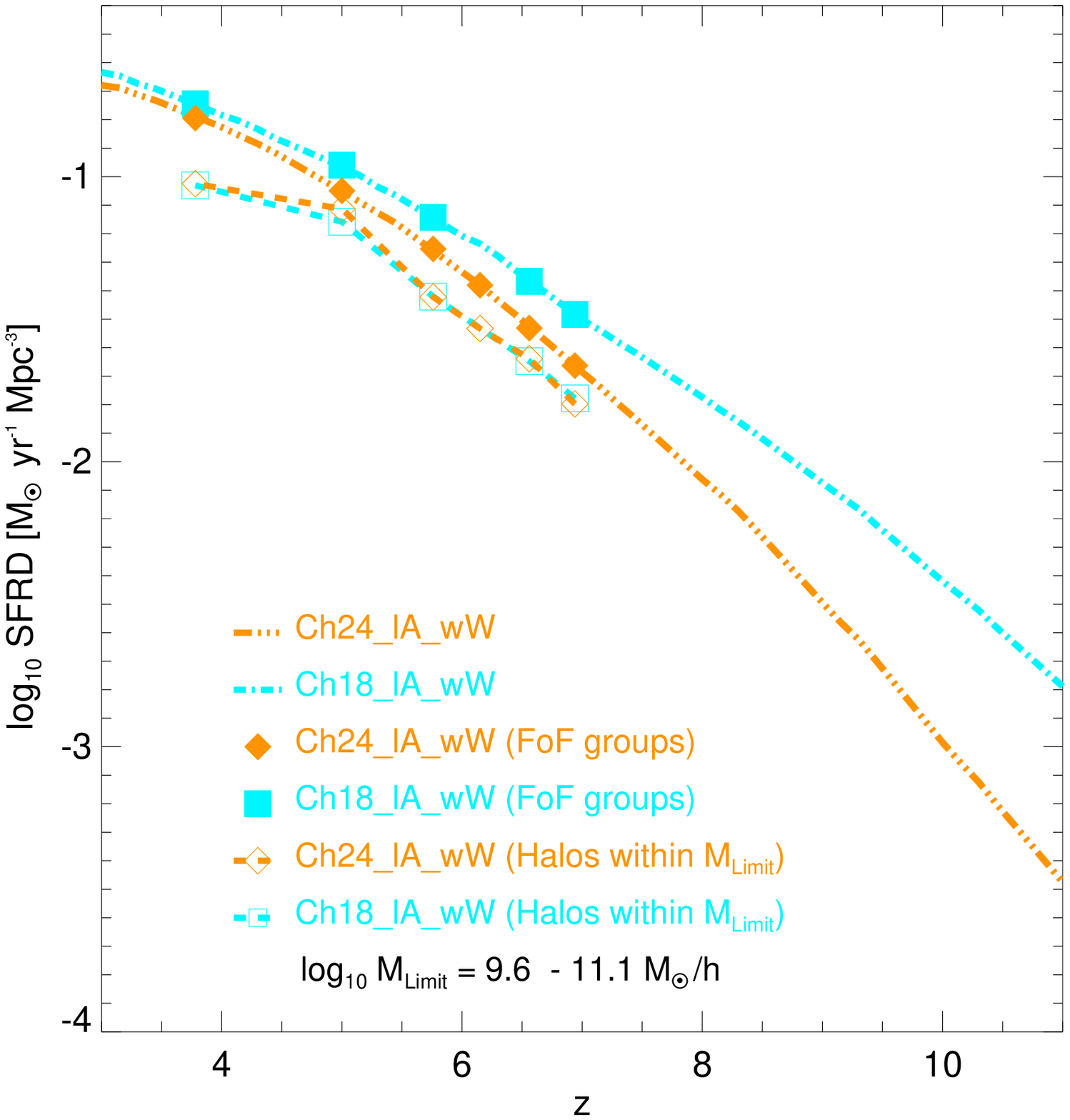}
\caption{Evolution of the cosmic star formation rate density: box size
  and resolution tests. {\textit{Top panel}}: we compare $a)$ the
  Chabrier IMF, Early AGN, Strong Winds configuration for two
  different box sizes: $L=24$ Mpc/$h$ (\textit{Ch24\tu eA\tu sW} - red
  triple dot-dashed line) and $L=12$ Mpc/$h$ (\textit{Ch12\tu eA\tu
    sW} - black dotted line) and $b)$ the Chabrier IMF, Late AGN, Weak
  Winds configuration for $L=24$ Mpc/$h$ (\textit{Ch24\tu lA\tu wW} -
  orange triple dot-dashed line) and $L=18$ Mpc/$h$ (\textit{Ch18\tu
    lA\tu wW} - cyan dot-dashed line). {\textit{Bottom left panel}}:
  the dotted black line represents the \textit{Ch12\tu eA\tu sW} run
  and the red triple dot-dashed line represents the \textit{Ch24\tu
    eA\tu sW} run, as in the top panel. At any given redshift, the
  filled black squares and red diamonds mark the total SFR density in
  collapsed structures. The open black squares and red diamonds show
  the SFR density in halos of mass $10^{9.6}$ M$_{\rm \odot}/h\le {\rm
    M}\le 10^{10.7}$ M$_{\rm \odot}/h$. {\textit{Bottom right panel}}:
  same as the bottom left panel for runs \textit{Ch18\tu lA\tu wW}
  (cyan dot-dashed line) and \textit{Ch24\tu lA\tu wW} (orange triple
  dot-dashed line). The open cyan squares and orange diamonds show the
  SFR density in halos of mass $10^{9.6}$ M$_{\rm \odot}/h\le {\rm M}
  \le 10^{11.1}$ M$_{\rm \odot}/h$.}
\label{fig_CSFRD_res}
\end{figure*}

In Figure \ref{fig_CSFRD_res} we compare the evolution of the cosmic
star formation rate density for runs with box size equal to $L=24$
Mpc/$h$, $L=18$ Mpc/$h$ and $L=12$ Mpc/$h$. In the top panel, four
simulations are considered: two in the late AGN $+$ weak Winds
scenario (\textit{Ch24\tu lA\tu wW} - orange triple dot-dashed line
and \textit{Ch18\tu lA\tu wW} - cyan dot-dashed line) and two in the
early AGN $+$ strong Winds scenario (\textit{Ch24\tu eA\tu sW} - red
triple dot-dashed line and \textit{Ch12\tu eA\tu sW} - black dotted
line). We see from the plot that runs with the same configuration
converge at redshift $z\sim4.5$. Moreover, the two simulations with
higher resolutions (\textit{Ch12} and \textit{Ch18}) show a higher
star formation rate density with respect to the other two runs because
they can resolve higher densities at earlier times.

In the bottom left panel the red triple dot-dashed and black dotted
lines refer, respectively, to the \textit{Ch24\tu eA\tu sW} and the
\textit{Ch12\tu eA\tu sW} runs already shown in the top panel. The
filled black squares and red diamonds mark the cosmic SFR density in
collapsed FoF halos, at $4\le z\le7$. The plot shows that all the star
formation occurs inside halos, as we expect. Most importantly, the
open black squares and red diamonds mark the SFR in halos of mass
$10^{9.6}$ M$_{\rm \odot}/h\le {\rm M}\le 10^{10.7}$ M$_{\rm
  \odot}/h$. The lower limit corresponds to the mass of a halo
resolved with 100 DM particles in the \textit{Ch24} simulation (the
run with lower resolution). This is our mass confidence
limit\footnote{In the FoF algorithm every bounded structure formed of
  at least 32 particles is considered a halo. To avoid numerical
  spurious effects, in our analysis we consider only halos formed of
  at least 100 DM particles.}. The upper limit is the mass of the most
massive halo in the \textit{Ch12} simulation (the run with smaller box
size). In this mass range the two simulations converge at all the
redshifts considered in this work. We define this mass interval as the
``overlapping mass range''.

In the bottom right panel, we compare the \textit{Ch24\tu lA\tu wW}
(orange triple dot-dashed line) and the \textit{Ch18\tu lA\tu wW}
(cyan dot-dashed line) runs. The results of this panel are consistent
with the results shown in the bottom left panel. In this case, the
open cyan squares and orange diamonds mark the SFR in halos of mass
$10^{9.6}$ M$_{\rm \odot}/h\le {\rm M}\le 10^{11.1}$ M$_{\rm
  \odot}/h$. The upper limit is now higher than before, corresponding
to the larger box size of the \textit{Ch18} with respect to the
\textit{Ch12}. In this mass range the two simulations converge at all
the redshifts considered. This demonstrates that, even if the total
SFR in the different boxes does not converge untill $z\sim4.5$, the
SFR in collapsed structures with mass in the overlapping mass range
has converged at much earlier times.

\begin{figure*}
\centering 
\includegraphics[scale=0.85]{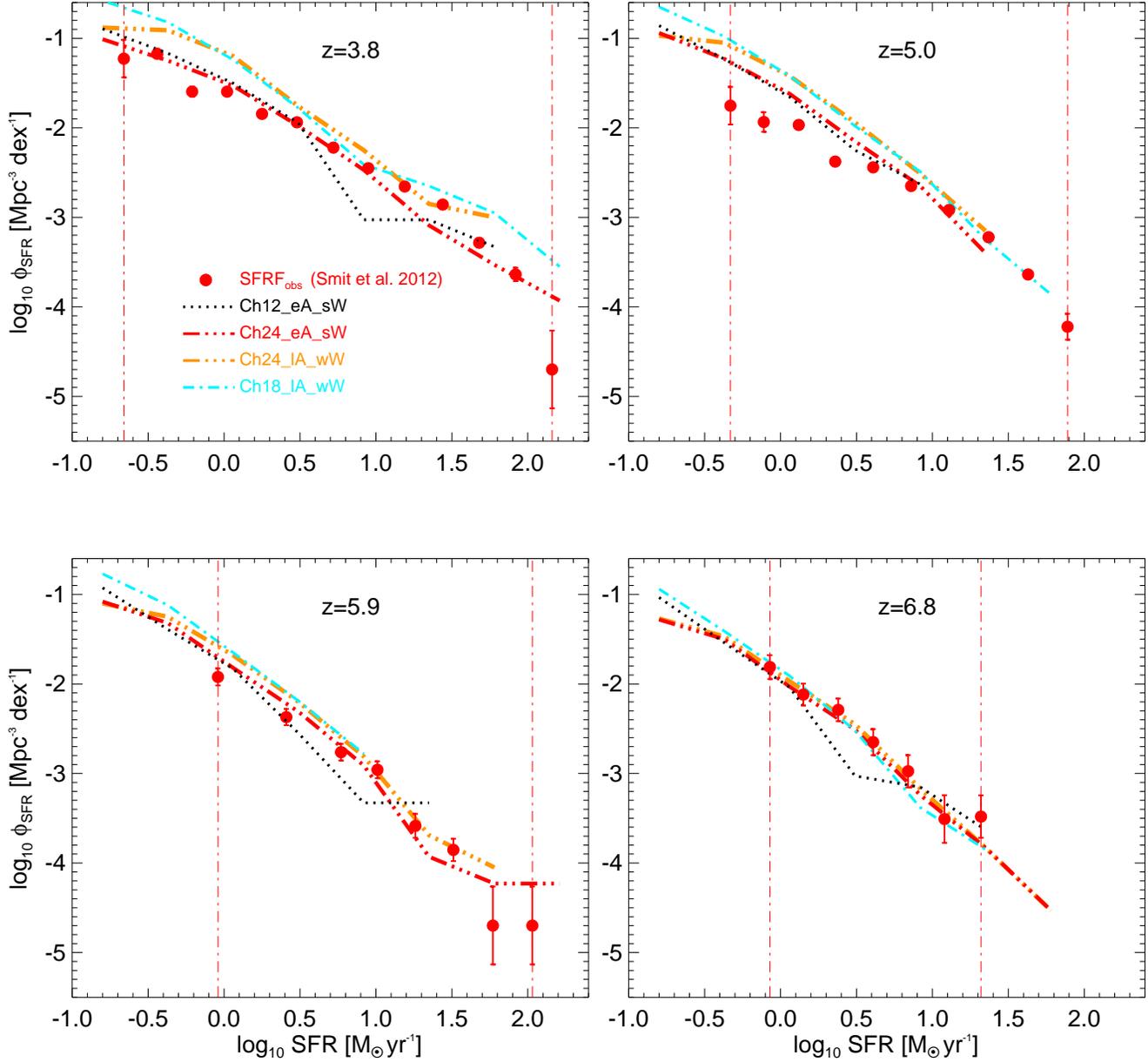}
\caption{Star formation rate functions at $z\sim4-7$: box size and
  resolution tests. At each redshift, the red filled circles with
  error bars and the vertical red dot-dashed lines represent the
  stepwise determinations of the SFR function and the observational
  limits of \citet{smit12}, respectively. Line-styles of different
  simulations are as in Figure \ref{fig_CSFRD_res}.}
\label{fig_SFRF_res}
\end{figure*}

In Figure \ref{fig_SFRF_res} we show the box size and resolution tests
for the star formation rate functions at $z\sim4-7$. We compare the
same simulations used above: \textit{Ch24\tu eA\tu sW} (red triple
dot-dashed line), \textit{Ch12\tu eA\tu sW} (black dotted line),
\textit{Ch24\tu lA\tu wW} (orange triple dot-dashed line) and
\textit{Ch18\tu lA\tu wW} (cyan dot-dashed line). Overplotted are the
data (red filled circles with error bars) and the observational limits
(vertical red dot-dashed lines) of \citet{smit12}. At all redshifts
considered, the \textit{Ch12} run shows poorer statistics at high SFR
with respect to the other runs. This is due to its smaller box size,
since there is a positive SFR$-$halo mass correlation and the box size
sets an upper limit on the mass of the halos in the simulation. At
redshift $z\le5$ ($5<z\le7$) the \textit{Ch12} run converges with the
corresponding \textit{Ch24\tu eA\tu sW} run in the range
$\log\,\frac{{\rm SFR}}{{\rm M}_{\odot}\ {\rm yr}^{-1}}\apprle0.5$
($\log\,\frac{{\rm SFR}}{{\rm M}_{\odot}\ {\rm
    yr}^{-1}}\apprle0.2$). On the other hand, inside the observational
windows of \citet{smit12} the \textit{Ch18} run agrees well with the
corresponding \textit{Ch24\tu lA\tu wW} at all redshifts. A small
difference is visible at $z=3.8$ for $\log\,\frac{{\rm SFR}}{{\rm
    M}_{\odot}\ {\rm yr}^{-1}}\apprle-0.3$. This is due to the fact
that in this SFR range the resolution limit of the \textit{Ch24}
simulations is reached. In fact, the \textit{Ch24\tu eA\tu sW} and the
\textit{Ch24\tu lA\tu wW} have the same SFRF value in the first SFR
bin (at all the redshifts considered), even if their configurations
are quite different. This feature is also visible in Figure
\ref{fig_SFRF} where the various \textit{Ch24} runs are compared.

To summarise, all the box size and resolution tests presented above
show that, even if the cosmic SFR density converges at $z\sim4.5$, our
results are robust at $z\le7$, provided only halos of mass ${\rm M}
\ge10^{9.6}$ M$_{\rm \odot}/h$ are taken into account.

\label{lastpage}
\end{document}